\documentclass{elsart}
\usepackage{epsfig,graphicx}
\usepackage{amsfonts}
\usepackage{amssymb}
\usepackage{amsmath}

\def\rootfig{}
\begin{document}

\begin{frontmatter}



\title{Multistable Solitons in the Cubic-Quintic Discrete Nonlinear
Schr\"odinger Equation}


\author[sdsu]{R.\ Carretero-Gonz\'alez\corauthref{cor1}},
\ead[url]{http://www.rohan.sdsu.edu/$\sim$carreter}
\corauth[cor1]{Corresponding author}
\author[sdsu]{J.D.\ Talley},
\author[sdsu]{C.\ Chong},
\author[israel]{B.A.\ Malomed}

\address[sdsu]
{
Nonlinear Dynamical Systems Group\thanksref{nlds},
Computational Science Research Center\\ and
Department of Mathematics \& Statistics, \\
San Diego State University, San Diego, CA 92182-7720, USA}
\thanks[nlds]{{\em URL:} {\tt http://nlds.sdsu.edu}}

\address[israel]
{
Department of Interdisciplinary Studies, Faculty of Engineering,\\
Tel Aviv University, Tel Aviv 69978, Israel
}

\begin{abstract}
We analyze the existence and stability of localized solutions in
the one-dimensional discrete nonlinear Schr\"{o}dinger (DNLS)
equation with a combination of competing self-focusing cubic and
defocusing quintic onsite nonlinearities. We produce a stability
diagram for different families of soliton solutions, that suggests
the (co)existence of infinitely many branches of stable localized
solutions. Bifurcations which occur with the increase of the
coupling constant are studied in a numerical form. A
variational approximation is developed for accurate prediction of
the most fundamental and next-order solitons together with their
bifurcations. Salient properties of the model, which distinguish it
from the well-known cubic DNLS equation, are the existence of two
different types of symmetric solitons and stable
asymmetric soliton solutions that are found in narrow
regions of the parameter space. The asymmetric solutions appear
from and disappear back into the symmetric ones via loops of
forward and backward pitchfork bifurcations.
\end{abstract}

\date{Accepted in \emph{Physica D}}


\end{frontmatter}

\section{Introduction}

Discrete nonlinear Schr\"{o}dinger (DNLS) equations constitute an
important class of discrete lattice models that are of great
interest in their own right \cite{Panos}, and also find direct
applications to the description of arrays of waveguides in
nonlinear optics, as predicted in Ref.\ \cite{Demetri}, and for
the first time realized experimentally in Ref.\ \cite{Silberberg},
which used a set of parallel semiconductor waveguides made on a
common substrate (see review article  Ref.~\cite{Nature} and further
references therein). In optics, quasi-discrete waveguide arrays can
also be created as virtual photonic lattices in photorefractive
materials (see review Ref.~\cite{Moti} and references therein),
and can be approximately described by the DNLS equations. The waveguide
arrays may support both spatial solitons \cite{Silberberg,Nature}
and quasi-discrete spatiotemporal collapse \cite{Shimshon}.

Besides nonlinear optics, the DNLS model also describes a Bose-Einstein
condensate trapped in a strong optical lattice (sinusoidal potential acting
on atoms in the condensate), as predicted theoretically \cite{BEC} and
observed in the experiment \cite{BECexperiment} (see also the recent review
Ref.~\cite{Chaos}). Additionally, DNLS equations may be naturally derived, in the
rotating-phase approximation, from various nonlinear-lattice models that
give rise to discrete breathers (alias intrinsic localized modes), see
theoretical papers Refs.~ \cite{review} and \cite{PhysToday}, and first reports of
the experimental making of these breathers in Ref.~ \cite{Sievers}.

Properties of discrete solitons in the DNLS equation with the
simplest, cubic, nonlinearity have been studied in detail,
including three dimensional settings \cite{vortex3d}, and are now
well understood. These solitons were experimentally observed in
arrays of nonlinear optical waveguides
\cite{Silberberg,Nature}. They also correspond, in the DNLS
approximation, to the intrinsic localized modes in more
sophisticated dynamical lattices.

Nonlinear Schr\"{o}dinger (NLS) equations with more complex
nonlinearities were studied in detail in continuum models. As well
as their cubic counterparts, such models are of interest by
themselves, and may also have direct applications \cite{Dauxois}.
In particular, glasses and organic optical media whose dielectric
response features the cubic-quintic (CQ) nonlinearity, i.e., a
self-defocusing quintic correction to the self-focusing cubic Kerr
effect, are known \cite{CQoptical}. Properties of solitons in the
NLS equations with the CQ nonlinearity may be very different from
those in the simplest cubic equation, especially in the case when
the higher-order nonlinearity is combined with a periodic
potential. Recently, a great variety of multistable solitons with
different numbers of peaks and different symmetries (even, odd,
etc.) have been found in the CQ NLS equation embedded in the
linear potential of the Kronig-Penney (KP) type (a periodic array
of rectangular potential wells) \cite{BorisKP:05}, after bistable
solitons were studied in the CQ NLS equation with a single
rectangular potential well \cite{BorisKP:04}. Solitons in the
continuum cubic NLS equation with the KP potential have been
studied too \cite{SmerziKP}.

The limiting case of the CQ NLS equation with a very strong KP
potential naturally reduces to the DNLS equation with the CQ
nonlinearity, and our objective in this paper is to construct
solitons in this discrete model and explore their stability. The
model is not only interesting by itself (as is shown in the
present work), but may also be realized experimentally in the form
of an array of waveguides built of the above-mentioned optical
materials featuring the CQ nonlinearity \cite{CQoptical}. It is
relevant to mention that stable discrete solitons were recently
found in the DNLS equation with saturable nonlinearity
\cite{discrete-saturable,Samuelsen} (note that the latter model
was introduced back in 1975 by Vinetskii and Kukhtarev
\cite{Russian}), and, moreover, optical discrete solitons
supported by the saturable self-defocusing nonlinearity were
experimentally created using the photovoltaic effect in a
waveguiding lattice built into a photorefractive crystal
\cite{photovoltaic}. We show in this work that discrete solitons
in the CQ model are very different from their counterparts
investigated in the aforementioned works\
\cite{discrete-saturable,Samuelsen,photovoltaic}, (most
importantly, they feature a great multistability, as shown below)
due to the fundamental fact that, unlike the saturable
nonlinearity, the combination of the CQ terms features
\emph{competition} of self-focusing and defocusing types. For the
same reason, the solitons in the CQ DNLS equation are drastically
different from ones investigated earlier
\cite{higher-order,Weinstein} in the DNLS\ equation with a
\emph{single} nonlinear term of an arbitrary power (for instance,
quintic instead of cubic). Finally, a quantum version of a
finite-length DNLS equation (Bose-Hubbard model) with the CQ
nonlinearity and periodic boundary conditions was considered in
Ref. \cite{quantum}, where states with a small number $q$ of
quanta (in most cases, $q\leq 6$) were constructed.

The manuscript is organized as follows. In the next section, we introduce
the model and its stationary solutions, and derive a two-dimensional map to
generate discrete solitons corresponding to homoclinic solutions.
Section \ref{Sec:multistabilty} reports various (multistable) soliton
solutions and their stability. In Section \ref{Sec:bif}, we focus on
bifurcations that create/annihilate different solutions and account for
exchange of stability between them. In Section \ref{Sec:VA} we present an
analytical variational approximation that correctly predicts the main
bifurcation branches. Section \ref{Sec:conclusions} concludes the paper.

\section{The model and dynamical reductions \label{Sec:model}}

The one-dimensional cubic-quintic discrete nonlinear Schr\"{o}dinger (CQ
DNLS) equation is
\begin{equation}
i\dot{\psi}_{n}+C\Delta \psi _{n}+B|\psi _{n}|^{2}\psi _{n}-Q|\psi
_{n}|^{4}\psi _{n}=0,  \label{CQDNLS}
\end{equation}
where $\psi _{n}$ are the complex fields at site $n$ (in the case of the
above-mentioned array of optical waveguides, $\psi _{n}$ is
amplitude of the electromagnetic wave in the given core),
$\dot{\psi}\equiv d\psi /dt$ (in the above-mentioned model of the
waveguide array, the evolutional variable is actually not time but
the coordinate along the waveguide), and the discrete second
derivative (discrete-diffraction operator in the array of
waveguides) is $C\Delta \psi _{n}\equiv C\left( \psi _{n+1}+\psi
_{n-1}-2\psi _{n}\right) $, where $C$ is the constant of the
tunnel coupling between the cores. The third and fourth terms in
Eq.\ (\ref{CQDNLS}) represent, respectively, the cubic and quintic
nonlinearities. We assume $C,B,Q>0$, which (as said above)
corresponds to the most natural case of the self-focusing cubic
(Kerr) nonlinearity competing with its self-defocusing quintic
counterpart. Upon renormalization of $\psi $ and $t$, we set $B=2$
and $Q=1$.

Equation (\ref{CQDNLS}) conserves two dynamical invariants: the
Hamiltonian,\begin{equation} H=\sum_{n}\left[ C\psi _{n}^{\ast
}\left( \psi _{n+1}+\psi _{n-1}-2\psi _{n}\right) +\left\vert \psi
_{n}\right\vert ^{4}-\frac{1}{3}\left\vert \psi _{n}\right\vert
^{6}\right] ,
\end{equation}
and norm,
\begin{equation}
M=\sum_{n}\left\vert \psi _{n}\right\vert ^{2}  \label{M}
\end{equation}(in the application to the optical waveguide array, the latter is the total
power of the light signal).

\begin{figure}[th]
\centerline{ \epsfig{file=\rootfig 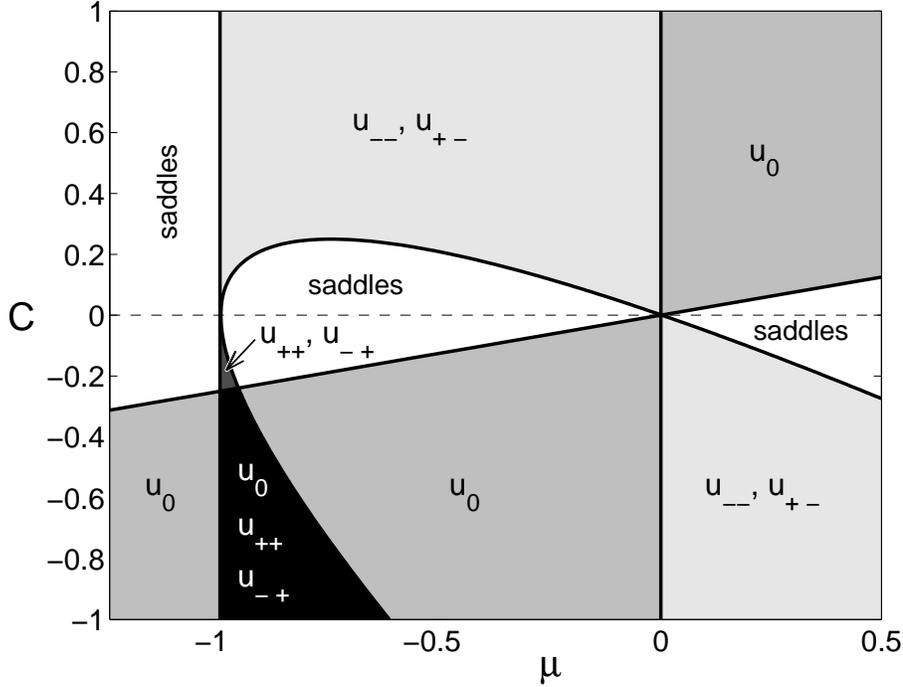,
width=12.0cm,angle=0,silent=} } \caption{Stability diagram for the
fixed points (FPs) of the two-dimensional map
(\protect\ref{2Dmap}). In non shaded areas all the FPs are
saddles. Shaded areas represent regions where the indicated FPs
are centers. The diagonal line and the tilted parabola correspond,
respectively, to $C=\protect\mu /4$ and $(C+1+\protect\mu
)^{2}=1+\protect\mu $. } \label{fixed_pt_stab.ps}
\end{figure}

\begin{figure}[th]
\centerline{ \epsfig{file=\rootfig 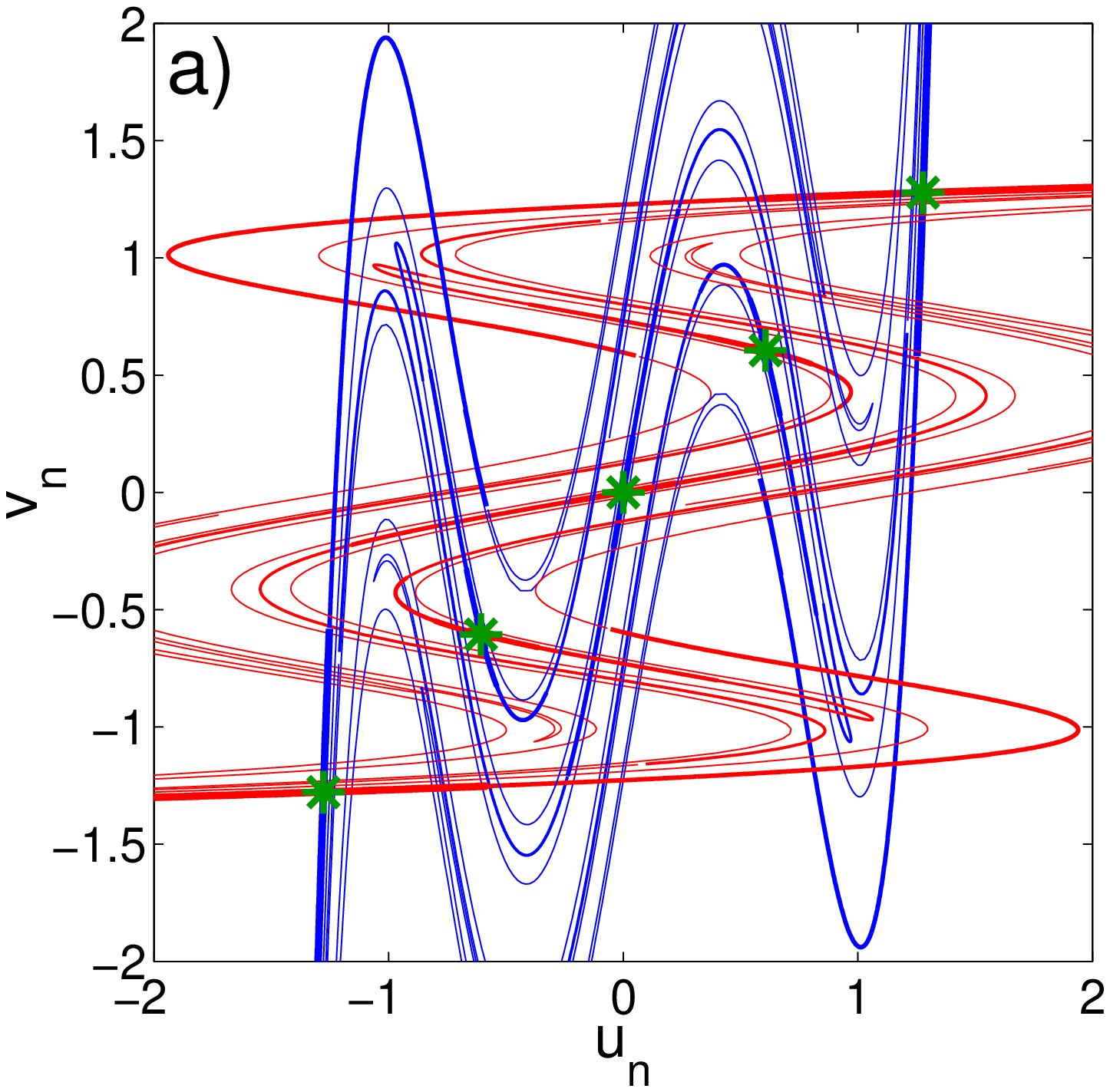,
width=5.3cm,angle=0,silent=} \epsfig{file=\rootfig
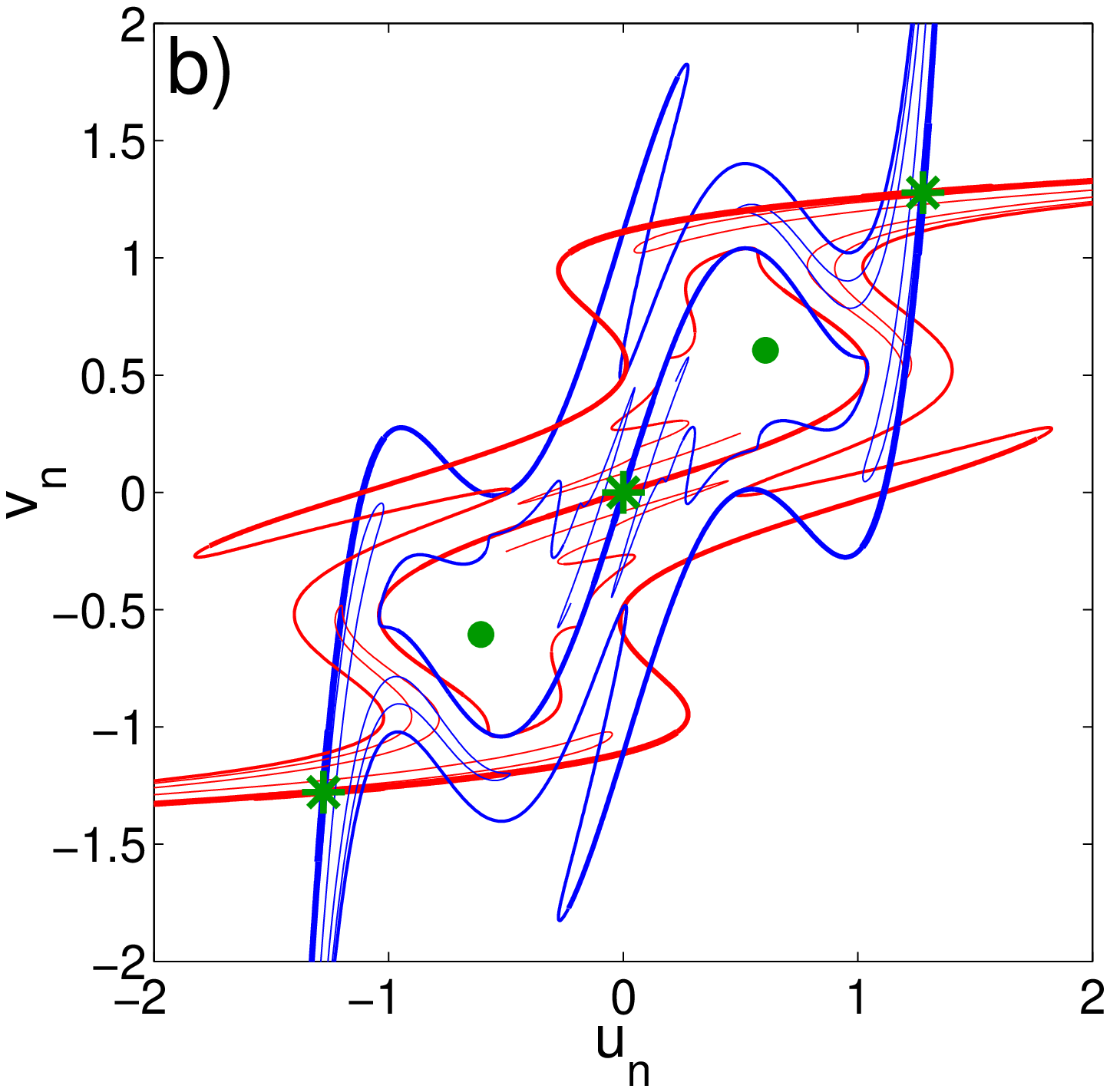, width=5.3cm,angle=0,silent=} } \centerline{
\epsfig{file=\rootfig 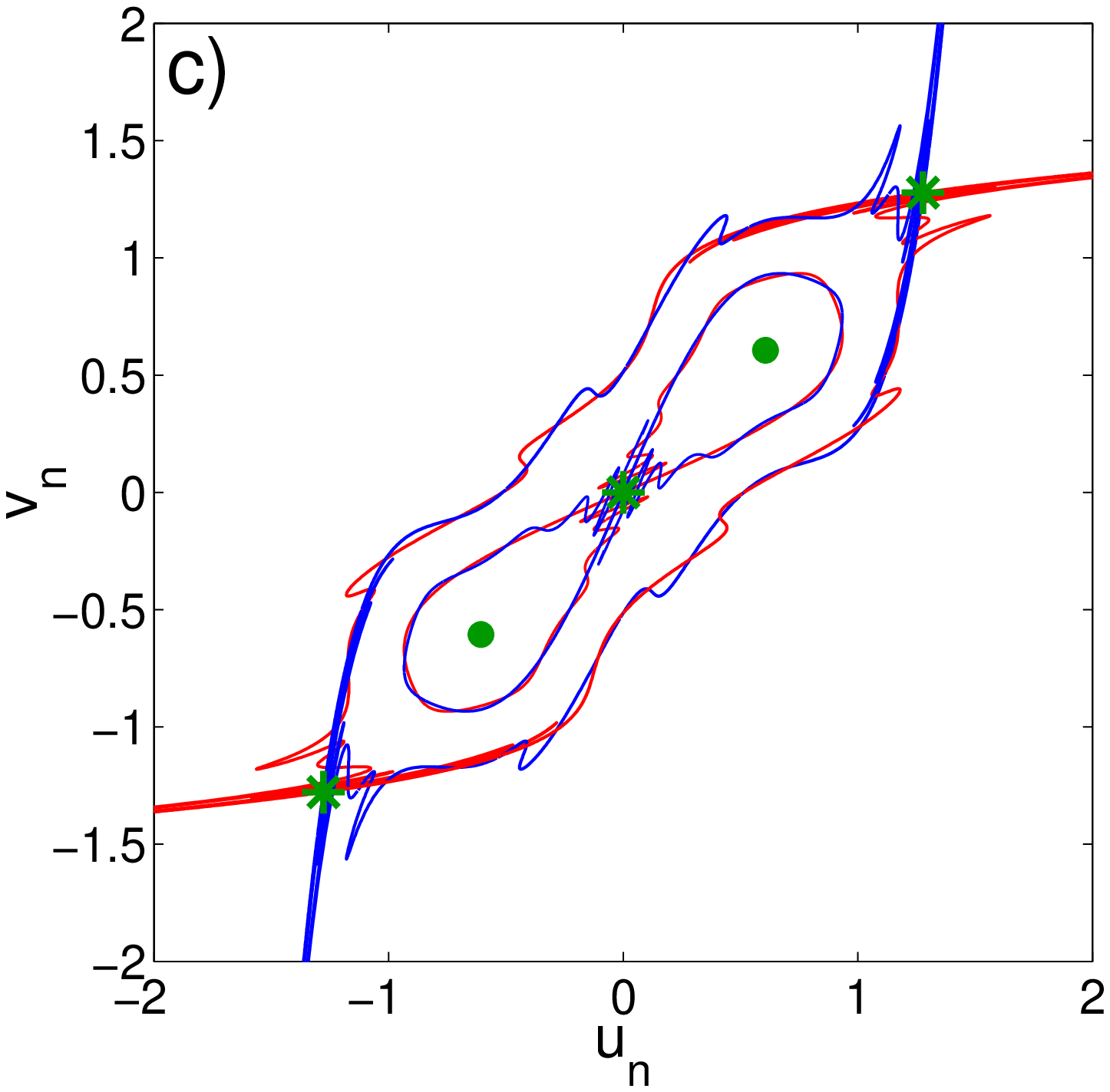,width=5.3cm
,angle=0,silent=} \epsfig{file=\rootfig
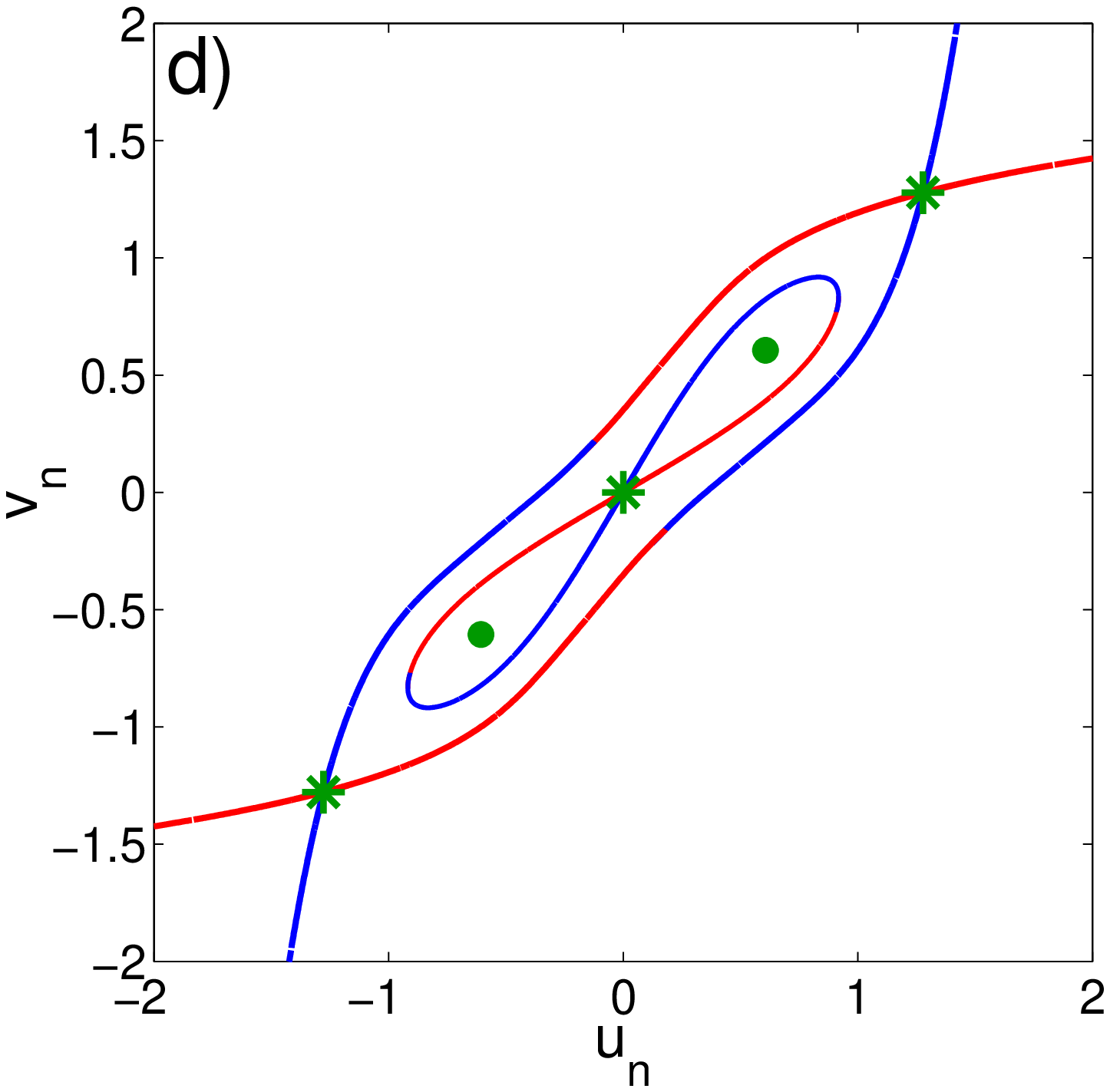,width=5.3cm ,angle=0,silent=} }
\caption{Homoclinic tangles generating localized solutions of Eq.\
(\protect \ref{CQDNLS}) for $\protect\mu =-0.6$, with the coupling
constant $C$ increasing from left to right and from top to bottom:
(a) $C=0.15$, (b) $C=0.4$, (c) $C=0.8$, and (d) $C=2$. Saddle
points and centers are designated by asterisks and circles,
respectively. For small $C$, the homoclinic intersections have a
much richer structure, including homoclinic and heteroclinic
connections between all five fixed points. For large $C$, only one
homoclinic and one heteroclinic solutions survive (together with
their $u_{n}\leftrightarrow -u_{n}$ symmetric counterparts). }
\label{manytangles.ps}
\end{figure}

We aim to look for a soliton solution with a frequency $\mu $ by
substituting
\begin{equation}
\psi _{n}=u_{n}\exp (-i\mu t)  \label{mu}
\end{equation}
in Eq. (\ref{CQDNLS}). The real stationary lattice field $u_{n}$ must solve
the equation
\begin{equation}
\mu u_{n}+C(u_{n+1}+u_{n-1}-2u_{n})+2u_{n}^{3}-u_{n}^{5}=0,
\label{CQDNLS-sta}
\end{equation}supplemented by the condition of vanishing of $u_{n}$ at
$n\rightarrow \pm \infty$, i.e., the soliton can be looked for
as a \textit{homoclinic solution} of
Eq.~(\ref{CQDNLS-sta}) (an alternative approach would be to use an algebraic
method of Ref.\ \cite{Tsironis}). Note that stationary soliton solutions of
Eq.~(\ref{CQDNLS-sta}) depend on two parameters, $\mu $ and $C$.

We will study soliton solutions and their stability by viewing
Eq.\ (\ref{CQDNLS-sta}) as a recurrence relation between
consecutive amplitudes, that can be cast in the form of a
two-dimensional map,
\begin{equation}
\left\{
\begin{array}{rcl}
u_{n+1} & = & \displaystyle au_{n}-v_{n}-2C^{-1}u_{n}^{3}+C^{-1}u_{n}^{5} \\[2ex]
v_{n+1} & = & u_{n},\end{array}\right.  \label{2Dmap}
\end{equation}
where $a\equiv 2-\mu /C$. Constant solutions to Eq.\ (\ref{CQDNLS})
correspond to fixed points (FPs) of this map. There exists at most
five FPs, that we arrange in increasing order and label as
$u_{-+},u_{--},u_{0},u_{+-},u_{++}$, with $u_{0}=0$ and
\begin{equation}
u_{\pm \pm }=\displaystyle\pm \sqrt{1\pm \sqrt{1+\mu }}.  \label{fpts}
\end{equation}
The stability of all the FPs within the framework of map (\ref{2Dmap}) can
be easily derived from the linearization of the map around these FPs,
leading to the stability chart displayed in Fig. \ref{fixed_pt_stab.ps}.

\begin{figure}[th]
\centerline{ \epsfig{file=\rootfig 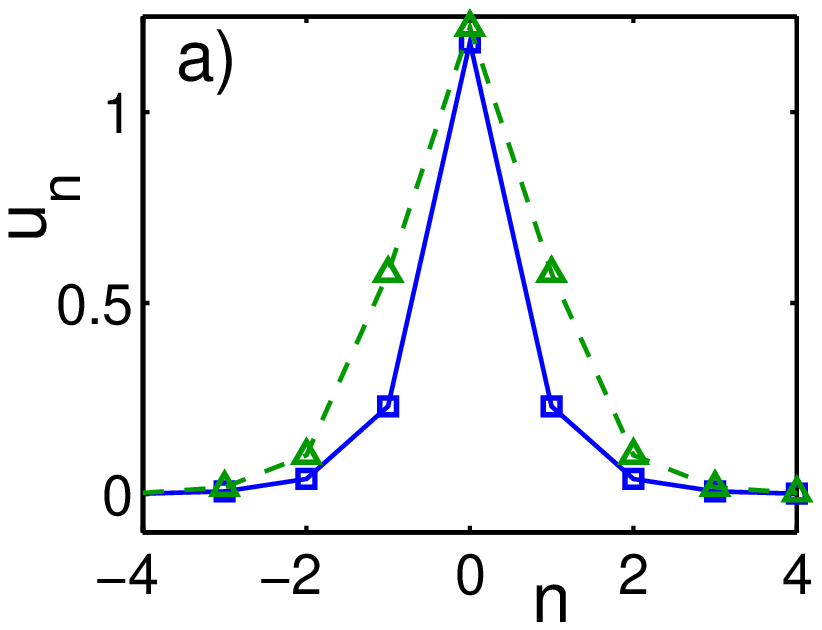,
width=5.3cm,angle=0,silent=} \epsfig{file=\rootfig
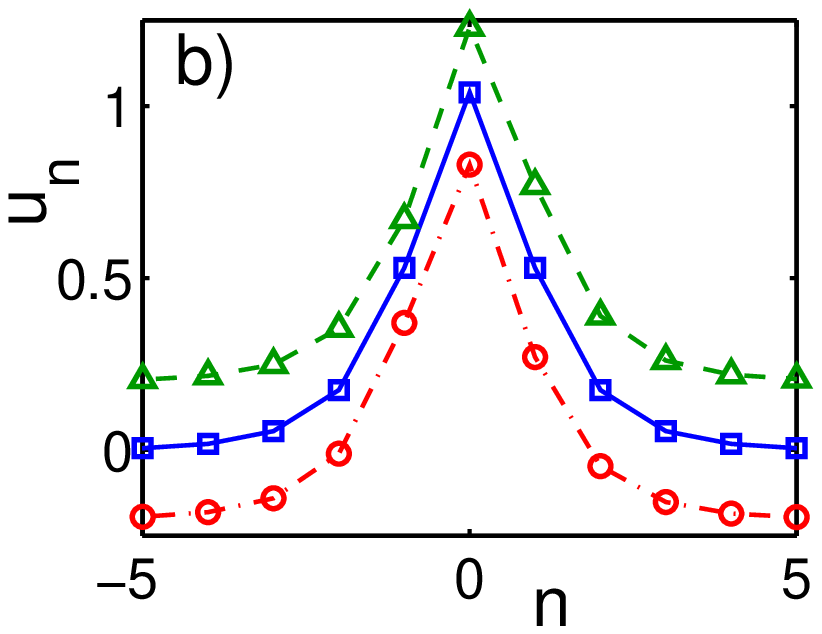, width=5.3cm,angle=0,silent=} } \centerline{
\epsfig{file=\rootfig 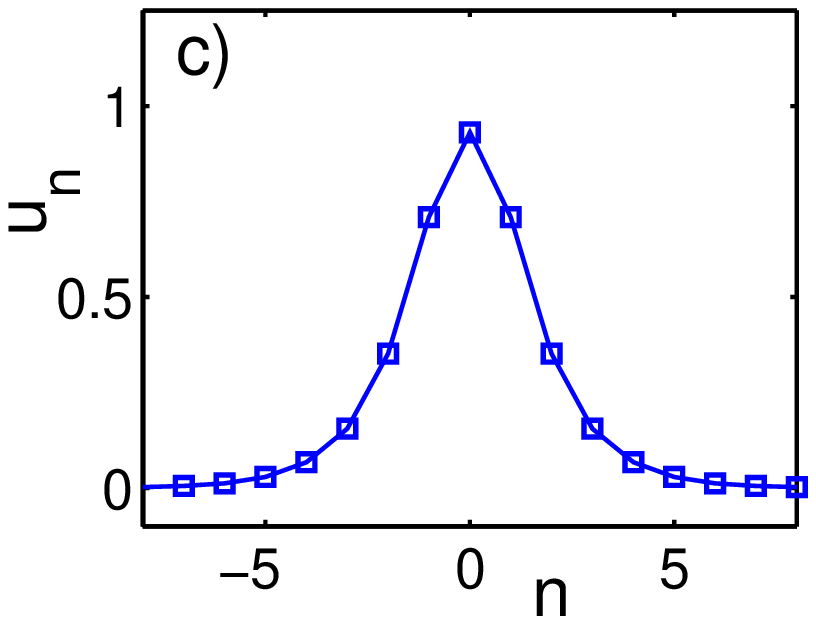,
width=5.3cm,angle=0,silent=} \epsfig{file=\rootfig
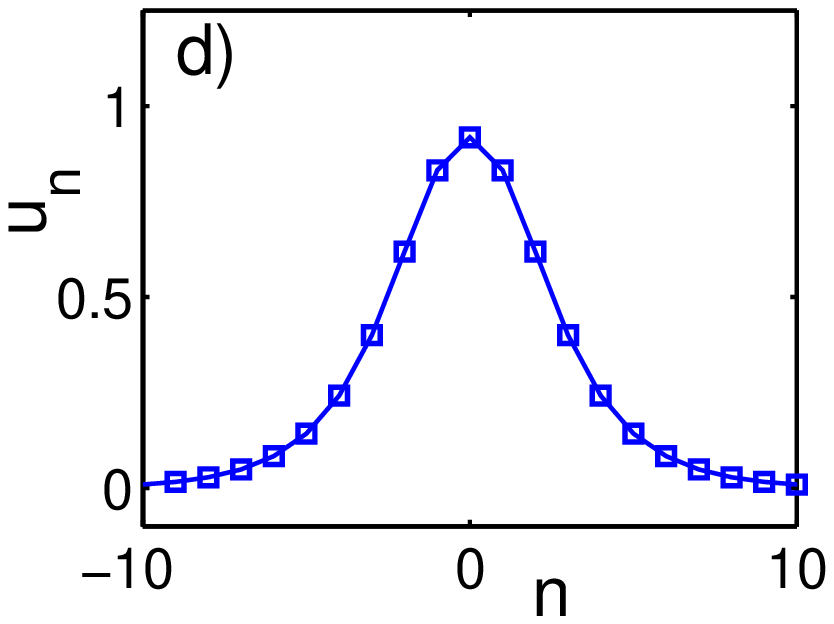, width=5.3cm,angle=0,silent=} } \caption{Typical
examples of soliton solutions generated by the homoclinic tangles
of Fig.\ \protect\ref{manytangles.ps} as the coupling parameter
$C$ increases. Insets pertain to the same values of parameters as
the corresponding insets in Fig.\ \protect\ref{manytangles.ps}.
(a) Pair of coexisting symmetric solution for small $C$. (b) As
$C$ increases, only one stable symmetric solution exists
(squares), together with a pair of asymmetric solutions (triangles
and circles) appearing in small regions of the parameter space
(the asymmetric solutions have been vertically displaced for
clarity of presentation). (c) Further increase in $C$ destroys all
stable solutions except for the symmetric one, that, as seen in
panel (d), tends towards the continuum soliton in the limit of
$C\rightarrow \infty $. The parameters are $\protect\mu =-0.6$ and
(a) $C=0.15$, (b) $C=0.4$, (c) $C=0.8$, (d) $C=2$.}
\label{manyorbits}
\end{figure}

In this framework, soliton solutions correspond to homoclinic orbits
connecting the FP at origin ($u_{0}=0$) with itself, in the case when it is
a saddle. The FP $u_{0}=0$ is a saddle for $\{\mu<0$ and $C>\mu/4\}$
and for $\{\mu>0$ and $C<\mu/4\}$. We are only interested in physically
meaningful couplings so we restrict our attention to $C>0$. Furthermore,
the regions for $\mu<-1$ and $\mu>0$ produce stable and unstable manifolds
that do not intersect each other. Therefore, in the remaining of this
work, we restrict our area of interest to $C>0$ with $-1<\mu<0$.

In Fig.\ \ref{manytangles.ps}, we depict a progression of the
homoclinic tangles (webs of such orbits) emanating from the saddle points as
the coupling constant $C$ increases. For small $C$, the
homoclinic-connection structure is very rich, including many orbits (i.e.,
many soliton solutions). As $C$ increases, many connections disappear
through a series of bifurcations (see below), so that a single homoclinic
loop survives at very large $C$, which corresponds to the well-known exact
soliton solution of the continuum CQ NLS equation \cite{Bulgaria}.

Note that, in contrast with the cubic DNLS equation, the CQ discrete
equation gives rise to a pair of extra fixed points that support
heteroclinic orbits (as shown in Fig.\ \ref{manytangles.ps}). The detailed
study of these heteroclinic orbits, which correspond to dark solitons or
kinks, falls outside the scope of this work. Here we concentrate on the
homoclinic orbits and (bright) soliton solutions corresponding to them.

In Fig.\ \ref{manyorbits} we depict a selection of typical solitons
corresponding to the homoclinic tangles in Fig.\ \ref{manytangles.ps}. Each
solution is numerically generated by taking the soliton shape as predicted
in an approximate form by the homoclinic intersections, and then applying a
Newton-type algorithm to find a numerically exact localized solution of Eq.\
(\ref{CQDNLS-sta}). In panel (a) we show a couple of solitons coexisting at
given parameter values. Panel (b) depicts a triplet of coexisting solutions
that form a part of a loop of pitchfork bifurcations responsible for the
creation of \emph{asymmetric} solutions (see below). Finally, panels (c) and
(d) show the unique site-centered (the maximum of the soliton is located at
a single central site, cf.\ Fig.\ \ref{pert_mu06_C01.ps}(a)) solution which
survives in the continuum limit, $C\rightarrow \infty $.

\section{Multistability of discrete solitons \label{Sec:multistabilty}}

\begin{figure}[th]
\centerline{
\epsfig{file=\rootfig 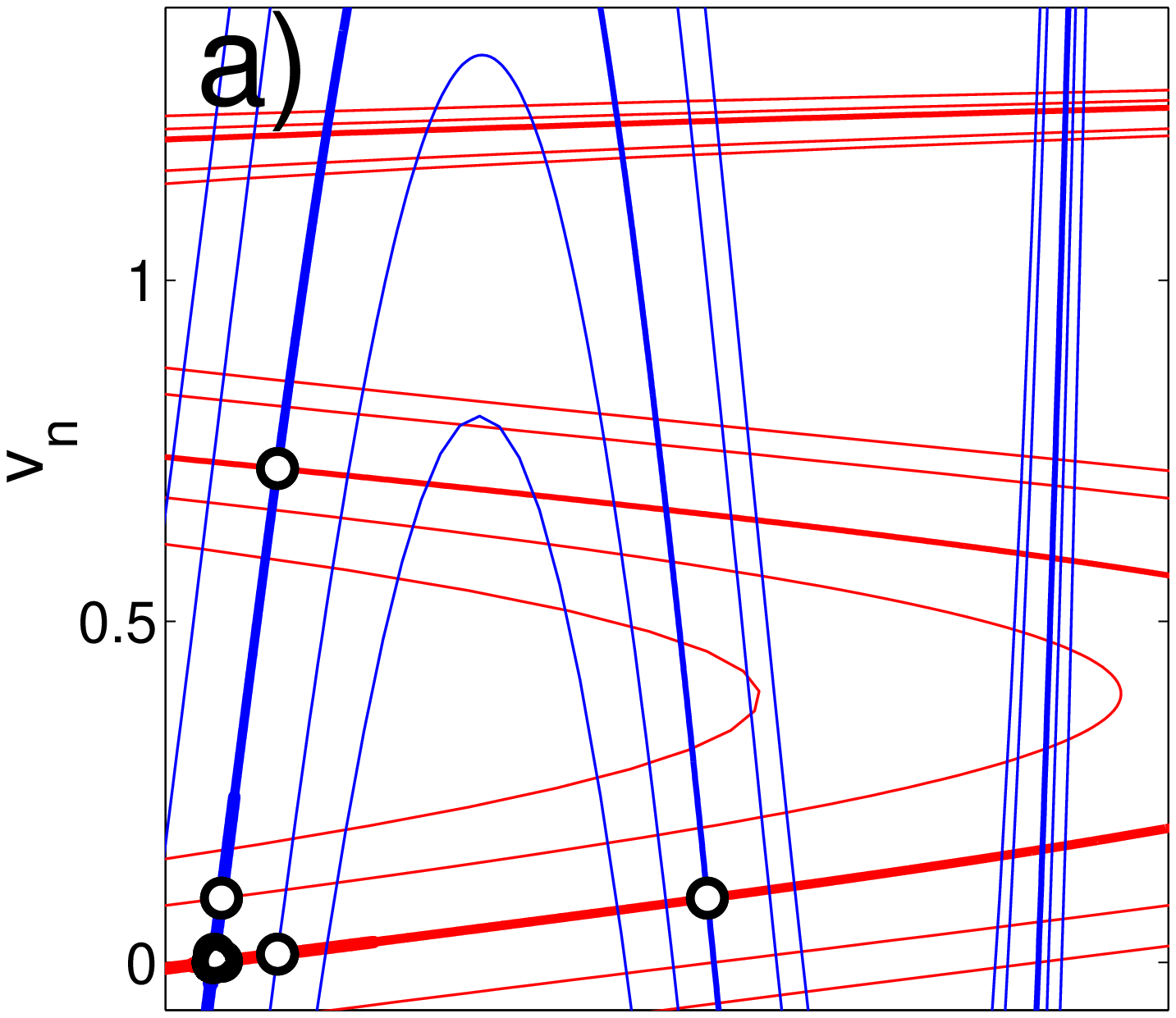,width=5.6cm,height=4.9cm ,angle=0,silent=}
\epsfig{file=\rootfig 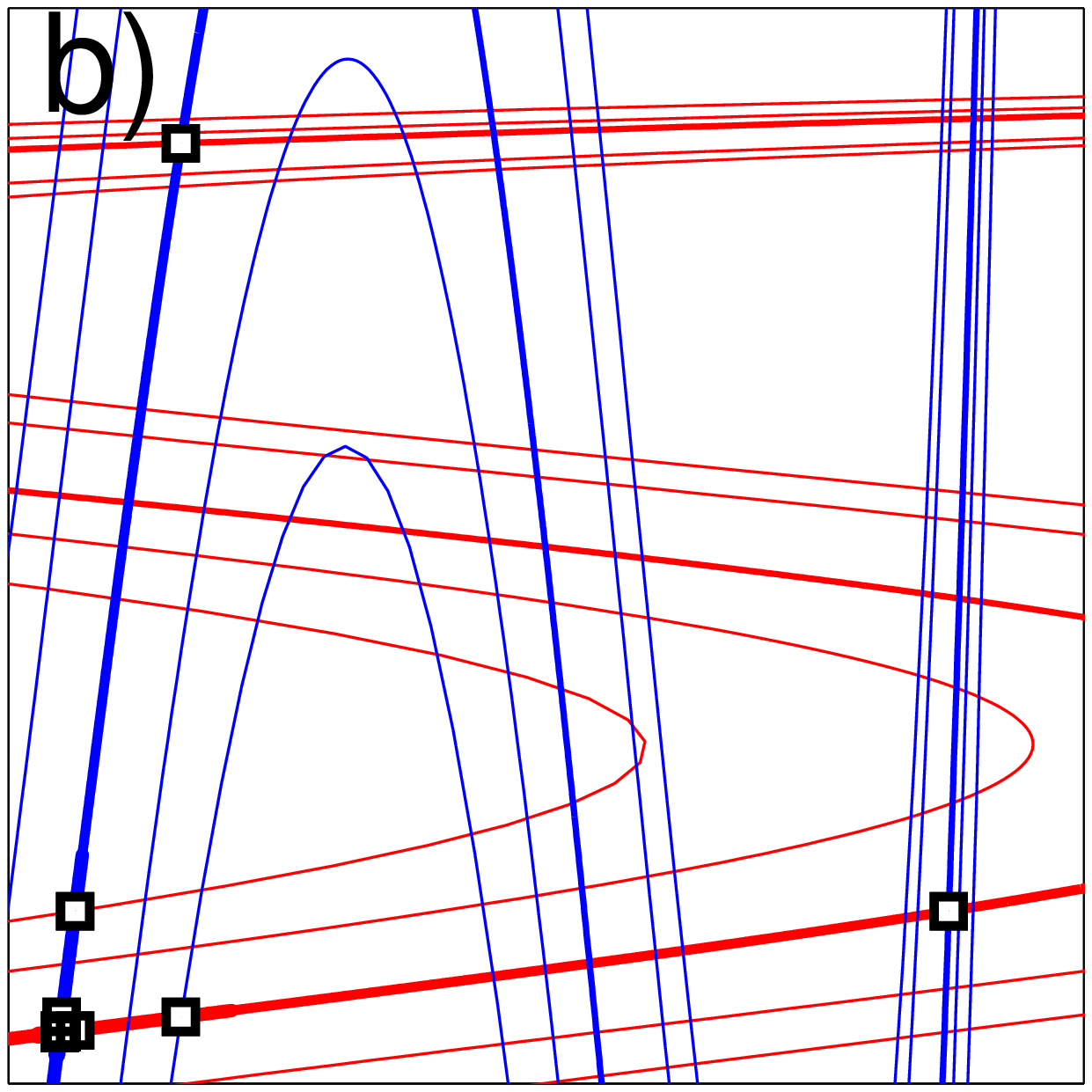,width=4.8cm,height=4.9cm,angle=0,silent=}
}
\centerline{
\epsfig{file=\rootfig 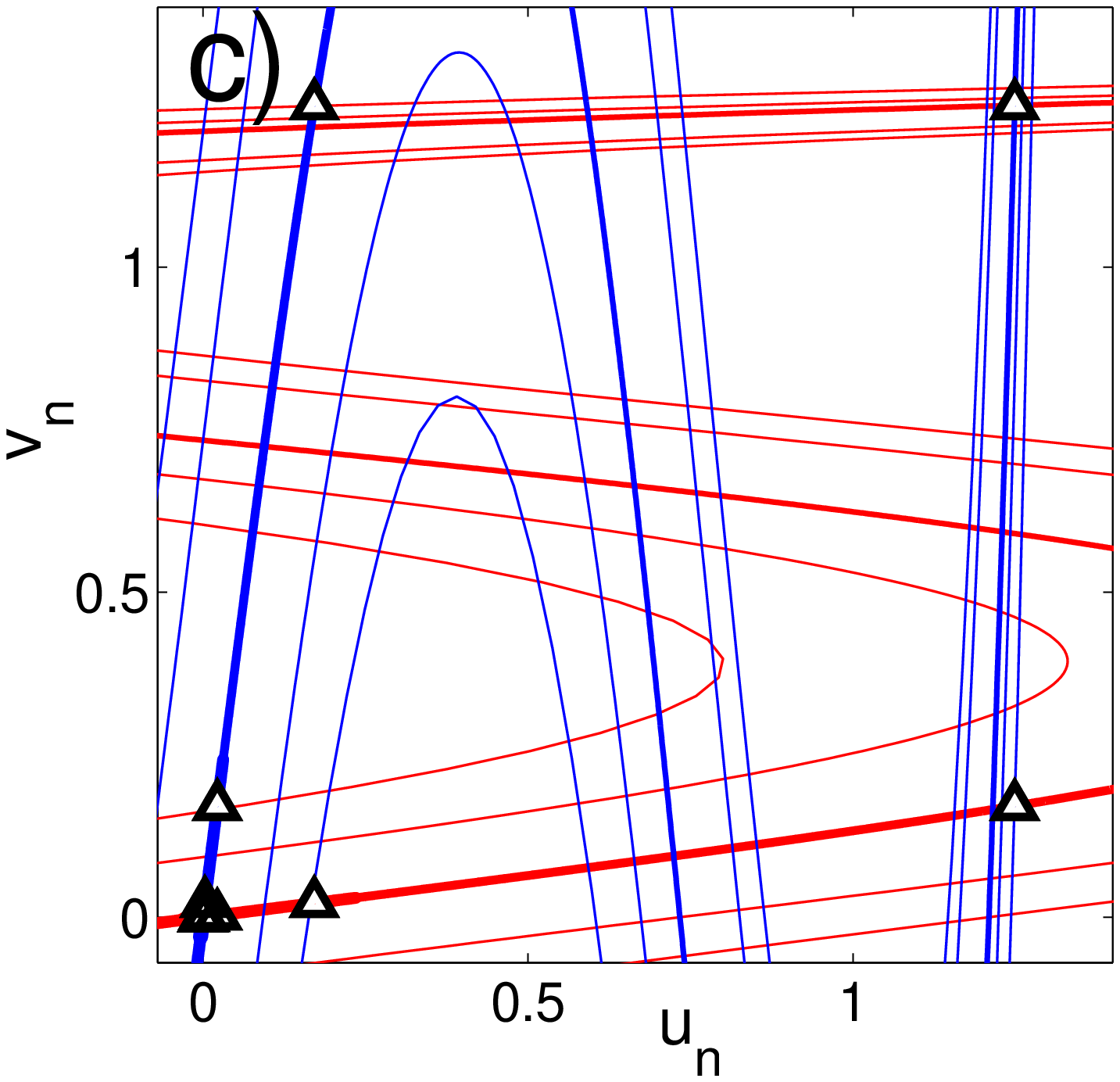,width=5.6cm,height=5.4cm ,angle=0,silent=}
\epsfig{file=\rootfig 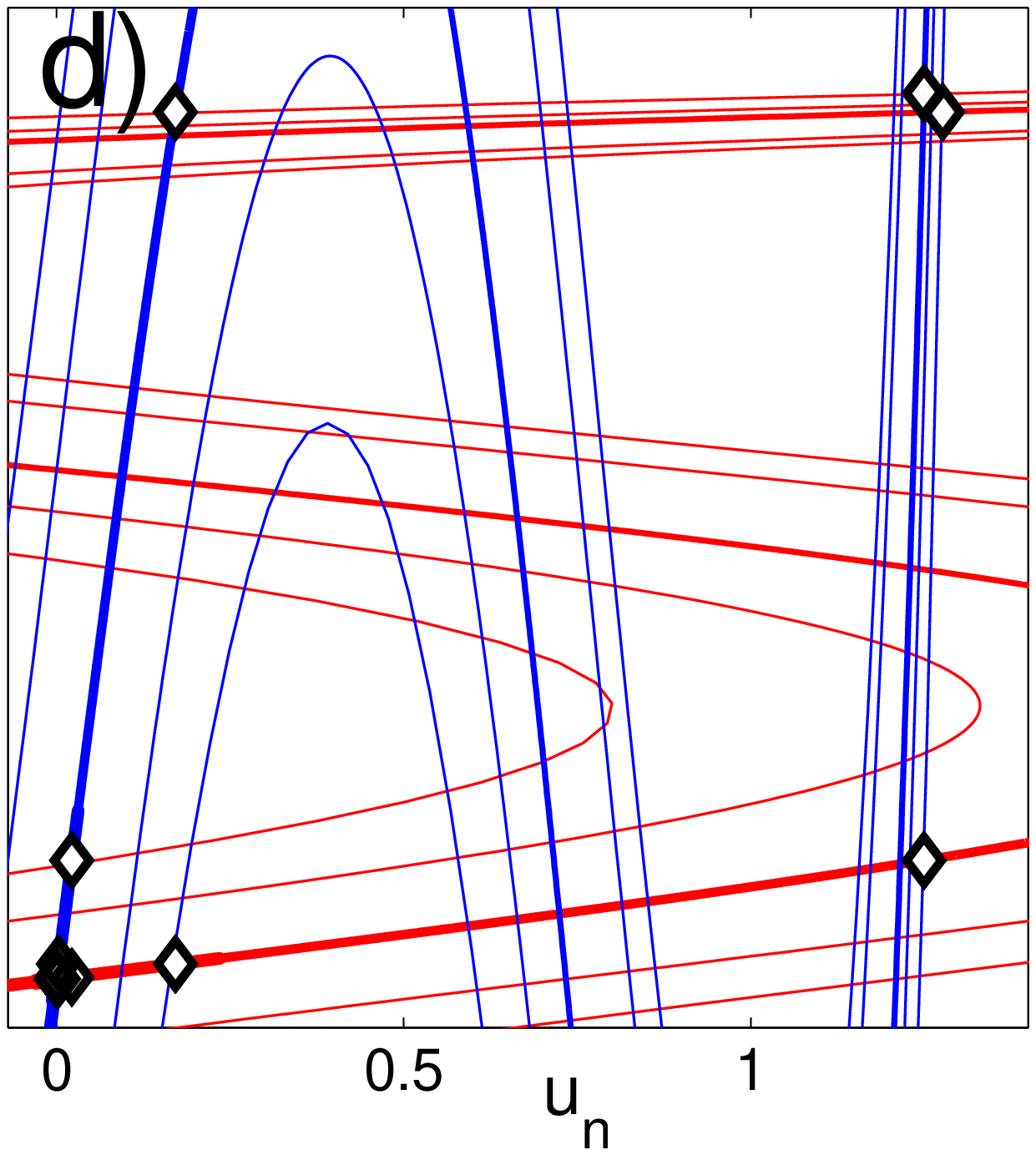,width=4.8cm,height=5.4cm,angle=0,silent=}
}
\caption{Multistability of discrete solitons in the CQ DNLS equation is
illustrated by displaying the homoclinic tangles generating \emph{stable}
localized solutions (see Fig.\ \protect\ref{pert_mu06_C01.ps}). Parameters
are $\protect\mu =-0.6$ and $C=0.1$ (which corresponds to the point marked
by the asterisk in the $T_{1}$ region in Fig.\ \protect\ref{stability.ps}).
Panels (a), (b), (c) and (d) correspond, respectively, to the discrete
solitons of the $S_{1}$, $T_{1}$, $T_{2}$, and $T_{3}$ types (see
definitions in the text).}
\label{homo_mu06_C01.ps}
\end{figure}

We now aim to explore the structure of the homoclinic tangles and
their bifurcations in detail, varying the parameters $\mu $ and
$C$. As previously mentioned, for small $C$ the rich homoclinic
structure leads to the coexistence of multiple solitons at the
same values of $\mu $ and $C$, which is a distinctive feature of
the CQ model with the competing nonlinearities: in the cubic DNLS
equations, this variety of solitons is not observed \cite{Panos}.
Principal types of the localized solutions for small $C$ (taking
$C=0.1$ as an example) are depicted in Fig.\
\ref{homo_mu06_C01.ps}. The following scheme is adopted to denote
different species of the solitons. The solution generated by the
first (main) homoclinic crossing of the stable and unstable
manifolds of the origin (see Fig.\ \ref{homo_mu06_C01.ps}(a)) is
denoted by $S_{1}$. This family corresponds to
site-centered solitons, the label
$S$ standing for \emph{short}, as the solitons of this type
correspond to the shortest family of homoclinic crossings. As $C$
is increased, the $S_{1}$ soliton suffers a series of alternating
stability switches (bifurcations). We use the notation $S_{k}$ to
denote solitons corresponding to the series of \emph{stable}
regions between the stability switches.
In this work we do not consider higher-order crossings corresponding to
repeated iterates in both the stable and unstable manifolds, that would
produce \emph{bound states} of the discrete solitons, alias multi-humped
or multibreather
(non fundamental) solutions \cite{ref:multibreathers}.

The solitons generated by the homoclinic crossings in Figs.\
\ref{homo_mu06_C01.ps}(b)-(d) correspond, in our notation, to
$T_{1}$, $T_{2}$ and $T_{3}$ solitons, $T$ standing for
\emph{tall} solitons. They are generated by the second family of
crossings, and are characterized by a higher soliton maximum, in
comparison with their $S$ counterparts.

\begin{figure}[th]
\centerline{
\epsfig{file=\rootfig 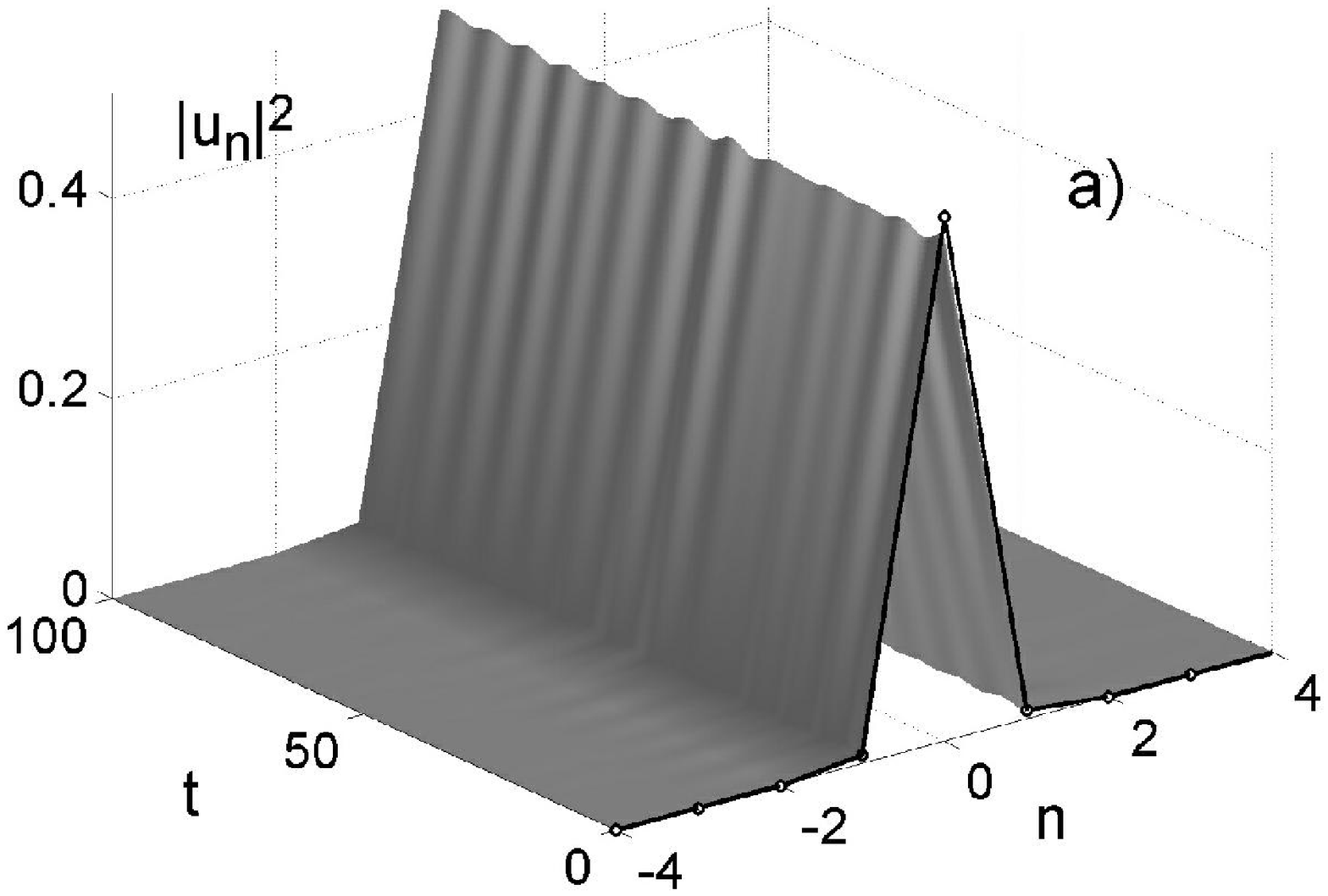,width=6.3cm,height=3.8cm ,angle=0,silent=}
\epsfig{file=\rootfig 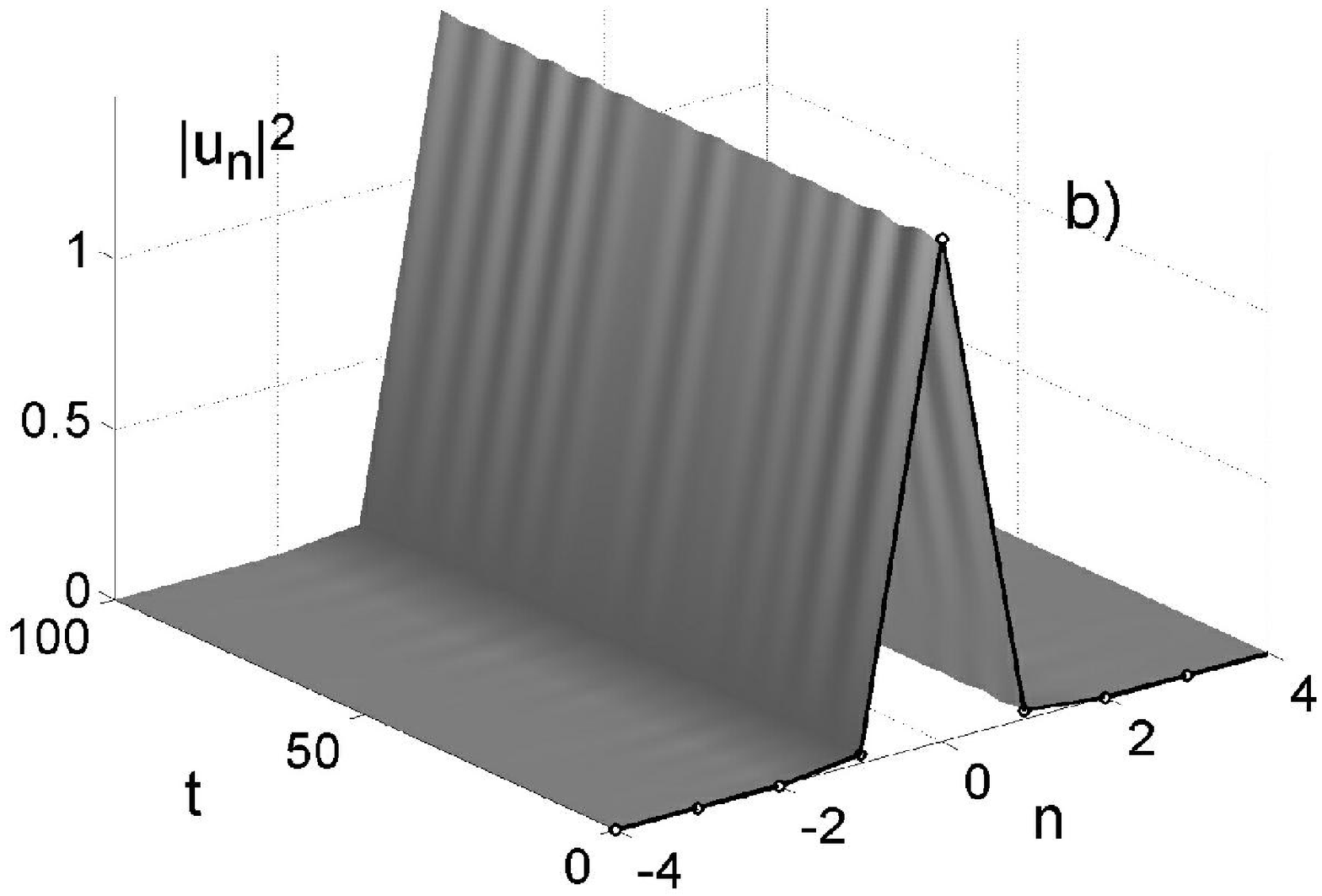,width=6.3cm,height=3.8cm ,angle=0,silent=}
}
\centerline{
\epsfig{file=\rootfig 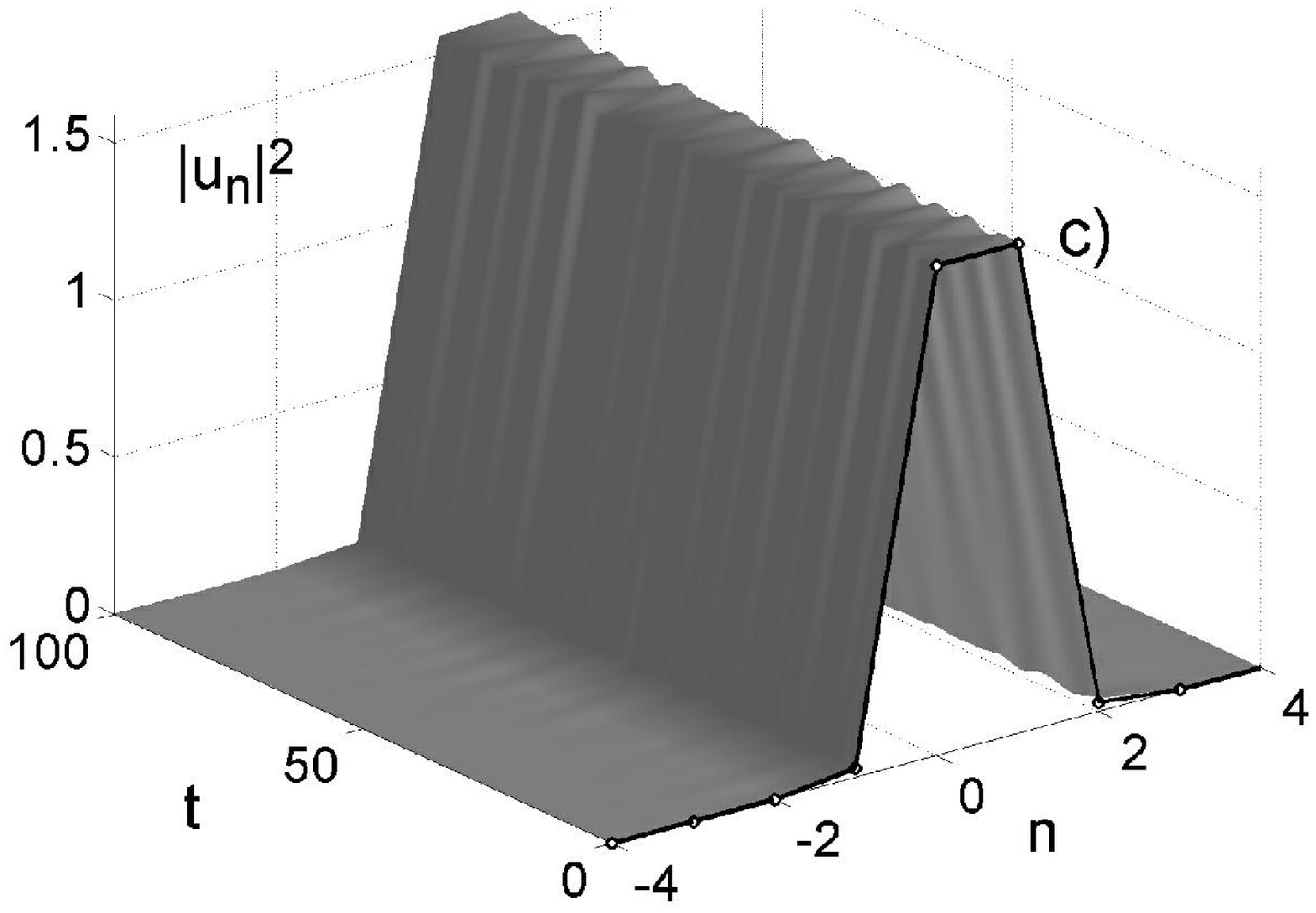, width=6.3cm,height=3.8cm,angle=0,silent=}
\epsfig{file=\rootfig 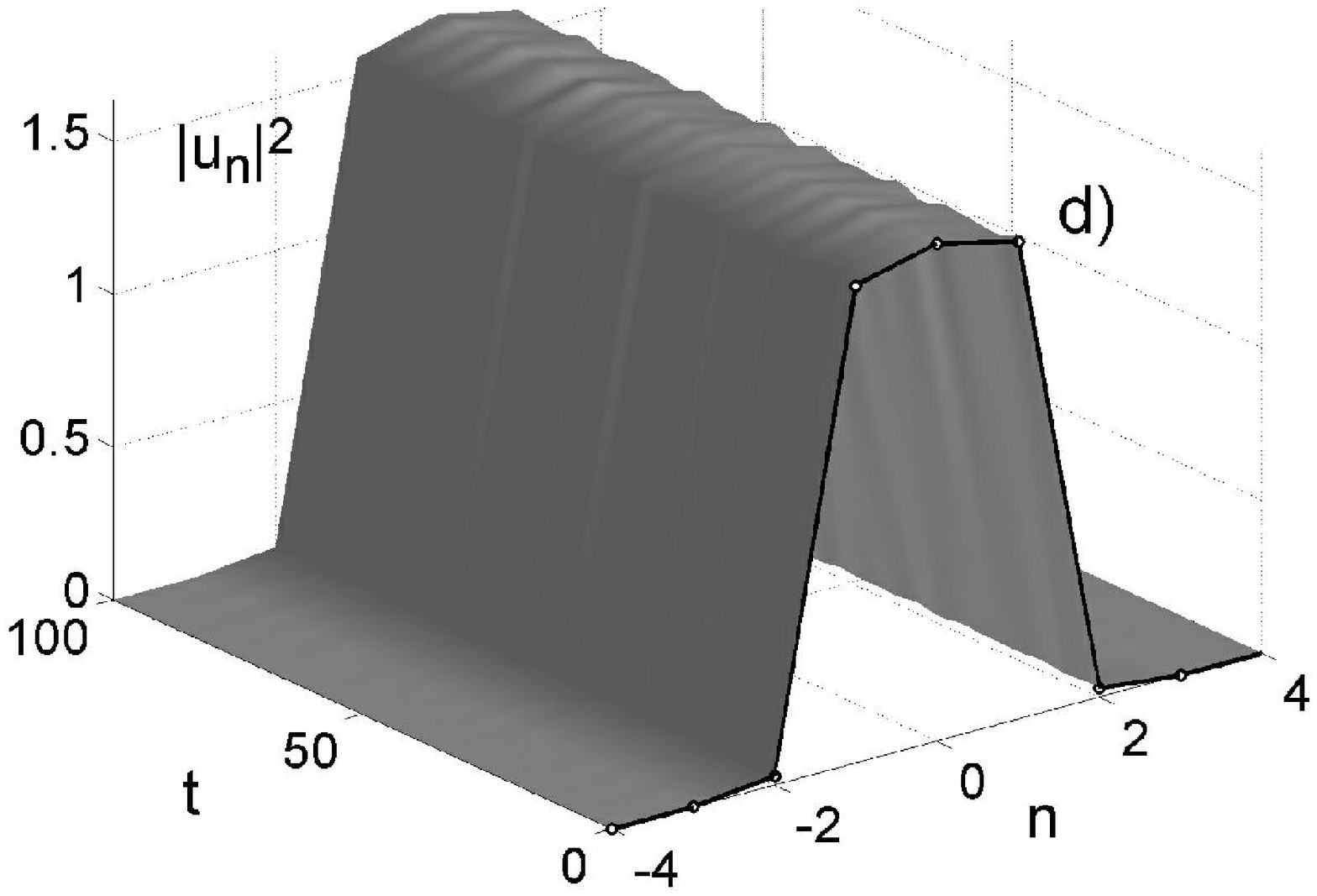, width=6.3cm,height=3.8cm ,angle=0,silent=}
}
\caption{Evolution of the solitons generated by the homoclinic intersections
depicted in Fig.\ \protect\ref{homo_mu06_C01.ps} after adding random
perturbations, with a relative amplitude of $5\%$, to the initial
configuration (dark line). }
\label{pert_mu06_C01.ps}
\end{figure}

The subscript in the notation for the $T_{k}$ and $S_{k}$ solutions also
helps to differentiate between \textit{site-centered} solutions, for $k$
odd, and \textit{bond-centered} (alias \textit{intersite-centered}) ones for
even $k$, which feature two central sites with equal magnitude, see\
examples in Fig.\ \ref{pert_mu06_C01.ps}.

The $S_{1}$, $T_{1}$, $T_{2}$ and $T_{3}$ solitons generated by the
highlighted crossings in Fig.\ \ref{homo_mu06_C01.ps} are all \emph{stable}
solutions. The stability was checked by calculating the respective
eigenvalues from Eq.\ (\ref{CQDNLS}) linearized around the stationary
solutions, and also verified by direct numerical integration of the full
equation\ (\ref{CQDNLS}) after adding a random perturbation to the soliton,
with a (rather large) relative amplitude of $5\%$. The evolution of the so
perturbed solutions is displayed in Fig.\ \ref{pert_mu06_C01.ps}, where the
unperturbed solitons are shown by dark lines for $t=0$. The perturbed
solutions oscillates about the unperturbed solitons, confirming their
stability (the oscillations do not fade because the system is conservative).

In order to identify a stability region of the soliton family in the $(\mu
,C)$ parameter plane, we start with a particular soliton solution of the $S$
or $T$ type, generated as described above, and then continued it by varying
the parameters, simultaneously computing the stability eigenvalues. The
continuation procedure started at small $C$ where, as said above, it is easy
to find a particular solution.

\begin{figure}[th]
\centerline{ \epsfig{file=\rootfig 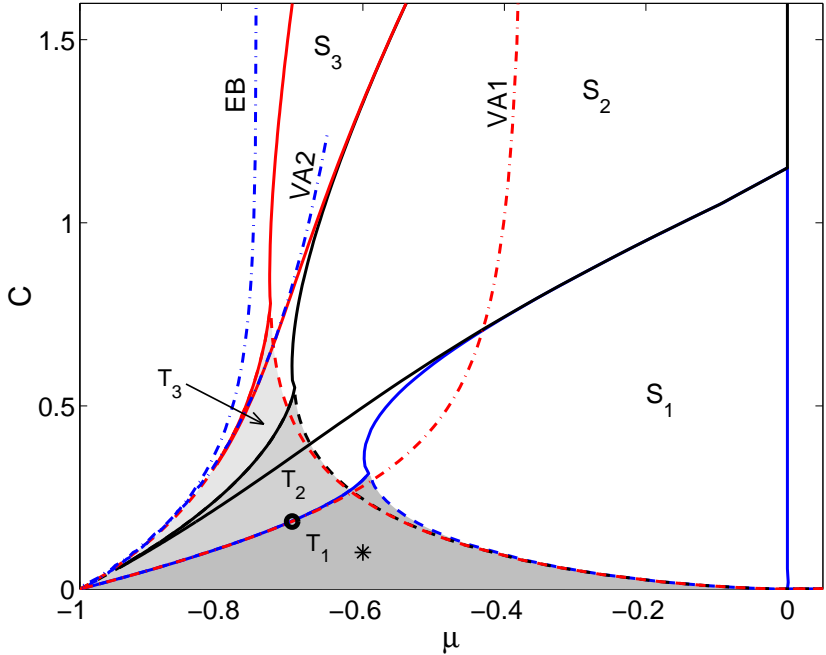, width=11cm
,angle=0,silent=} } \caption{The stability diagram for single-hump
localized solitons of the cubic-quintic discrete NLS equation
(\protect\ref{CQDNLS}). Note that each $S_{k}$ region has a
``wedge" that penetrates into the corresponding $T_{k}$ region.
The existence region of the solitons is bounded by the curve EB,
at which the homoclinic connections of the origin disappear. The
curves VA1 and VA2 show the variational approximations described
in Section \protect\ref{Sec:VA}. A more detailed description is
given in the text.} \label{stability.ps}
\end{figure}

The resulting stability diagram for the solitons of the
$S_{1,2,3}$ and $T_{1,2,3}$ types is displayed in Fig.\
\ref{stability.ps}. The $T_{k}$-stable regions (shaded in Fig.\
\ref{stability.ps}) feature a tent-like shape, with the base on
the line $C=0$ with $-1\leq \mu \leq 0$. Since all these regions
share the common base, there should be a nontrivial area where,
presumably, the $T_{k}$ solitons exist and are \emph{stable for
all values of} $k$. It is also apparent from the stability diagram
in Fig.\ \ref{stability.ps} that the stability regions for the
$S_{k}$ solutions with different $k$, in contrast to their $T_{k}$
counterparts, do not intersect each other. Therefore, the solitons
of the $S$ type feature no multistability.

We note that each $S_{k}$ stability region features a wedge that
penetrates into the region of stability of the $T_{k}$ solitons.
Particularly, the $T_{1}$ region is completely embedded in the
$S_{1}$ region. This property tends to suggest that, for any $k$,
there always exists non-empty regions where the $T_{k}$ and
$S_{k}$ solitons coexist and are \emph{simultaneously stable}. It
is interesting too that the $S$-stability regions outlive their
$T$-counterparts as $C$ increases. This is due to the fact that
the $T$ solutions correspond to the second family of homoclinic
crossings that disappear, in saddle-node bifurcations (see below),
earlier than the first family of the crossings (i.e., the $S$
solutions).

\begin{figure}[th]
\centerline{
\epsfig{file=\rootfig 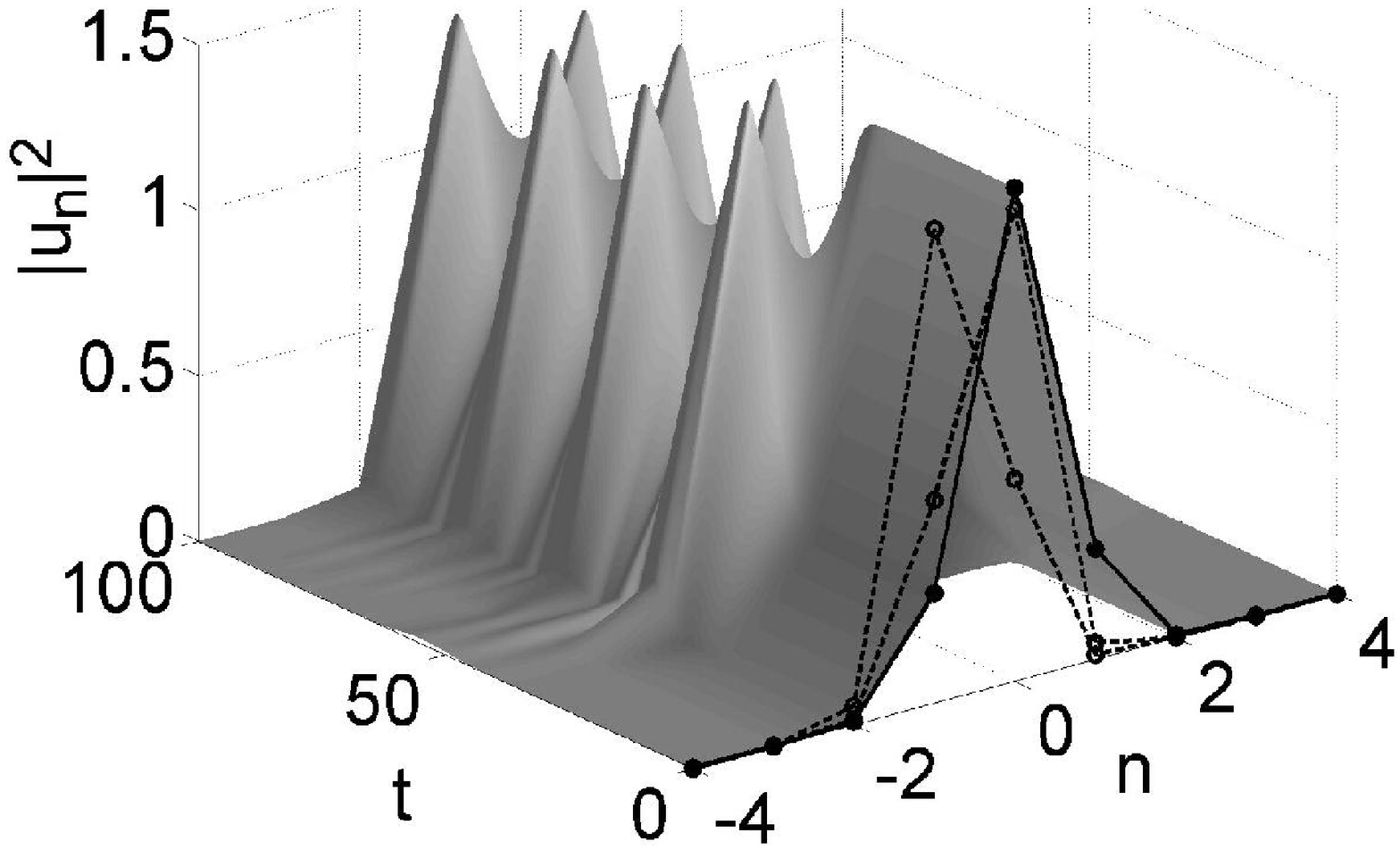, width=7.3cm,height=4.cm ,angle=0,silent=}
\epsfig{file=\rootfig 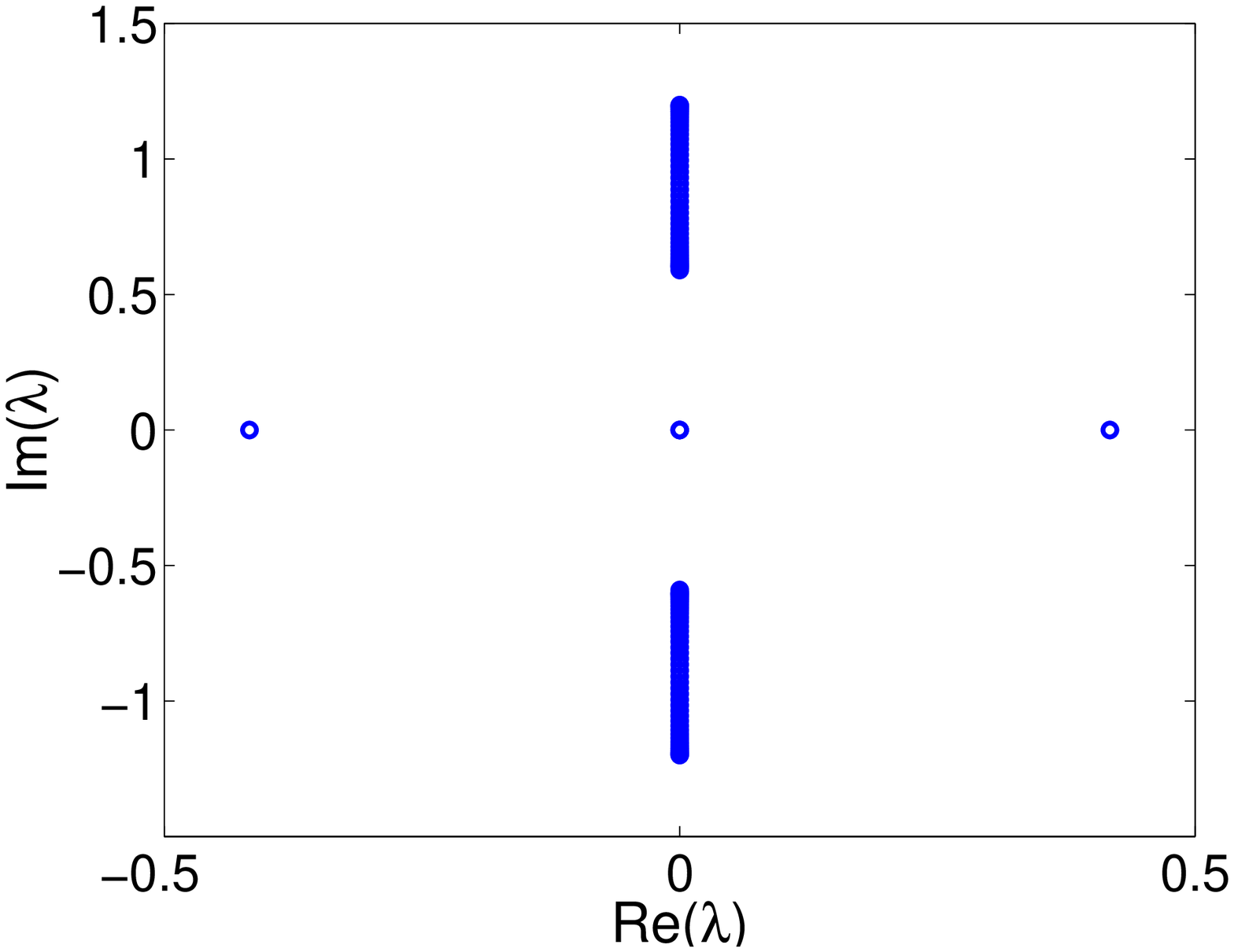, width=5.3cm,height=4.cm,angle=0,silent=}
} \medskip
\centerline{
\epsfig{file=\rootfig 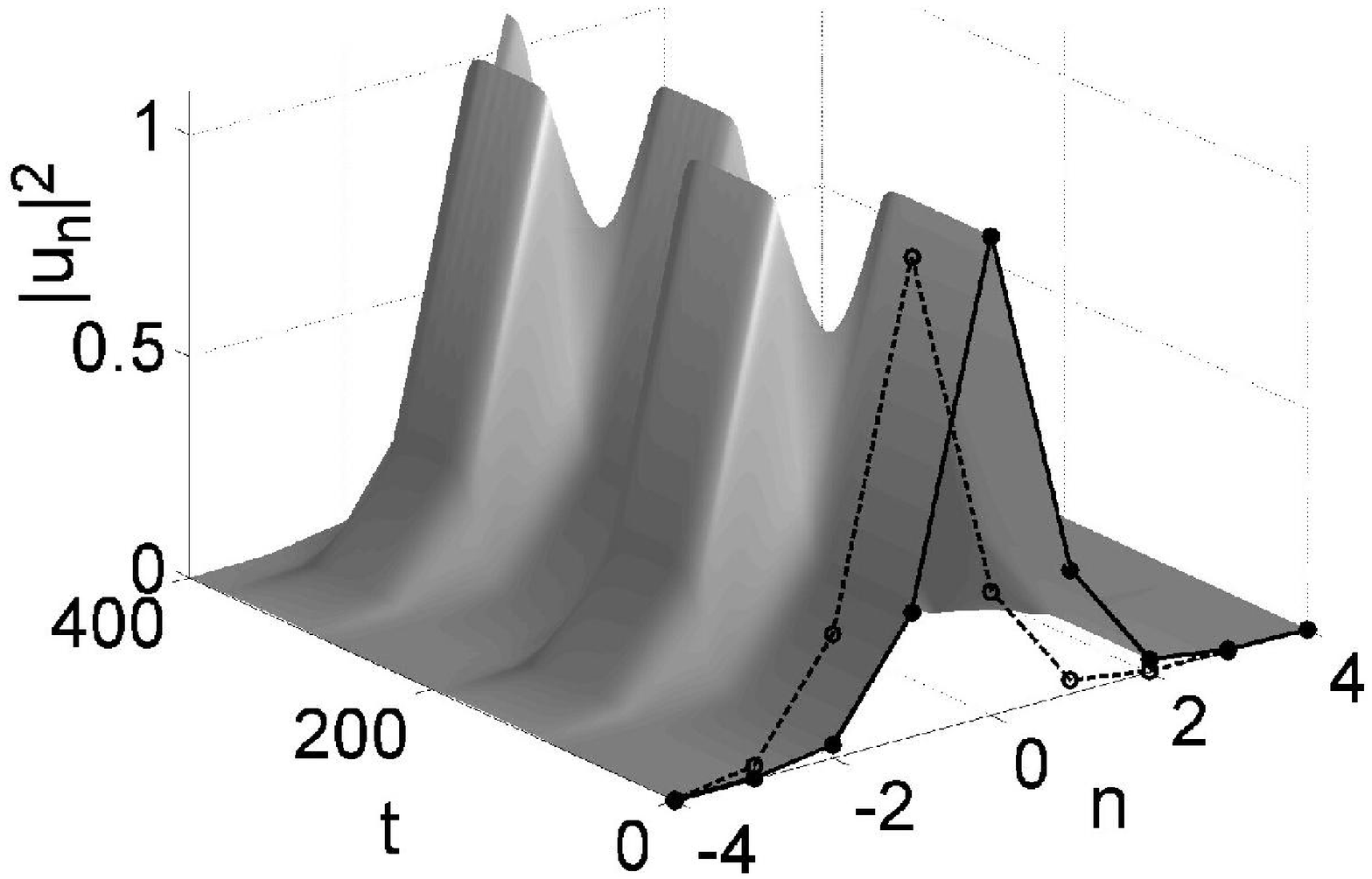, width=7.3cm,height=4.cm,angle=0,silent=}
\epsfig{file=\rootfig 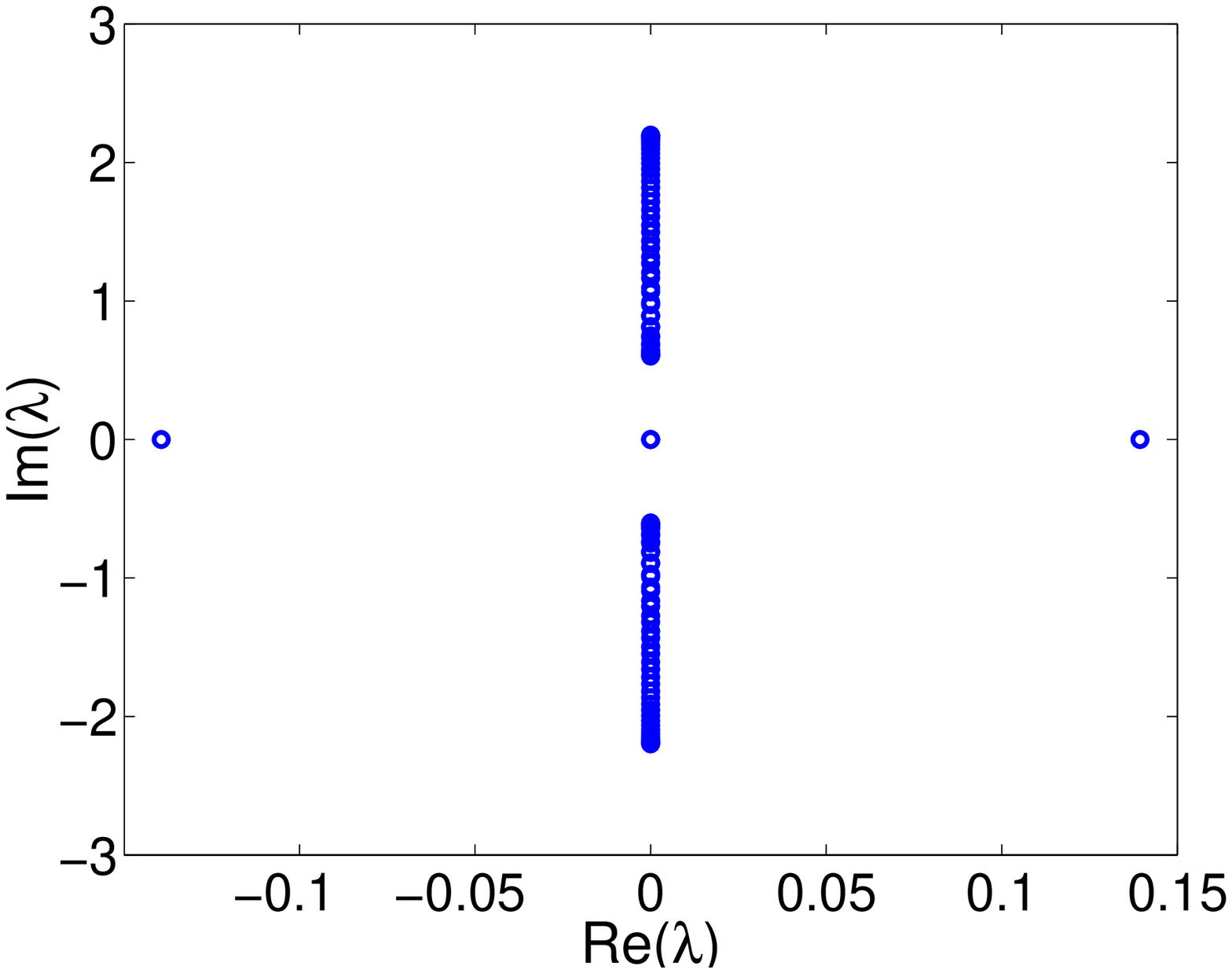, width=5.3cm,height=4.cm ,angle=0,silent=}
} 
\caption{Evolution of unstable solitons (left panels), together
with their instability spectrum (right panels). The top row
depicts the evolution of an unstable $S_{2}$ soliton, shown by the
dashed line in Fig.\ \protect\ref{manyorbits}(a) for $(\protect\mu
,C)=(-0.6,0.15)$, initiated by a small perturbation. After some
transient, the initial configuration (solid line) is changed by
one featuring oscillations between two asymmetric states (dashed
lines). The bottom row depicts the evolution of an unstable
symmetric $S_{1}$ solution from Fig.\ \protect\ref{manyorbits}(b)
for $(\protect\mu ,C)=(-0.6,0.4)$. The evolution amounts to
oscillations between the initial configuration (solid line) and
its translation by one site (dashed line). The respective
(in)stability spectra for the two solutions in the left column are
depicted in the right column. } \label{unstable_evol.ps}
\end{figure}

Existence regions for the solitons, which may be broader than the
regions where the solitons are stable, are defined as those in
which the stable and unstable manifolds emanating from the origin
do intersect. We have computed such a region numerically. In Fig.\
\ref{stability.ps}, it is located to the right of the
dashed-dotted line EB (\emph{existence boundary}). To the left of
this curve, no localized solutions are possible. As $C$ increases,
the EB curve approaches the line $\mu =-3/4$, that corresponds to
the existence border for solitons in the continuum limit of the CQ
equation \cite{BorisKP:05}. More specifically, the existence
regions of the $T$ solutions \emph{exactly coincide} with their
stability regions, i.e., the solitons of the\ $T$ type are
\emph{always stable}.
The existence region for the $S$ solutions has a complex structure, 
contained in the region where
homoclinic connections exists to the right of the EB curve, while the
stability area for these solitons is \emph{really smaller} than the
existence region. We also note
that (as explained in detail below) the $S_{k}$ solutions represent only two
distinct families of solitons, one site-centered (for odd $k$) and one
bond-centered (for even $k$).

Since some solitons of the $S$ type are unstable, it is necessary
to determine what the instability transforms them into. As an
example, in the top row of Fig.\ \ref{unstable_evol.ps} we display
the evolution of an unstable $S_{2}$ soliton in the stability
region of $T_{1}$ (for $(\mu ,C)=(-0.6,0.15)$, see asterisk in
Fig.\ \ref{stability.ps}), together with its instability
eigenvalues. The nonlinear evolution proceeds through periodic
oscillations between two asymmetric states (see dashed lines for
$t=0$).

In the bottom row of Fig.\ \ref{unstable_evol.ps} we show the evolution of
another unstable solution and its associated instability spectrum. In this
case, we take the $S_{1}$ solution at $(\mu ,C)=(-0.6,0.4)$, which is in the
gap between the stability regions of the $S_{1}$ and $S_{2}$ solutions. A
\emph{loop} of pitchfork bifurcations occurs in the gaps between consecutive
$S$-stability regions, see details in the next section. The evolution of
this unstable solution amounts to periodic oscillations between the original
$S_{1}$ soliton (the solid line at $t=0$) and its translation by one site
(the dashed line at $t=0$)). Therefore, this solution is a time-periodic
one, being close to a heteroclinic connection linking the two unstable
solutions, $S_{1}$ and its translation.
%

\begin{figure}[th]
\centerline{ \epsfig{file=\rootfig 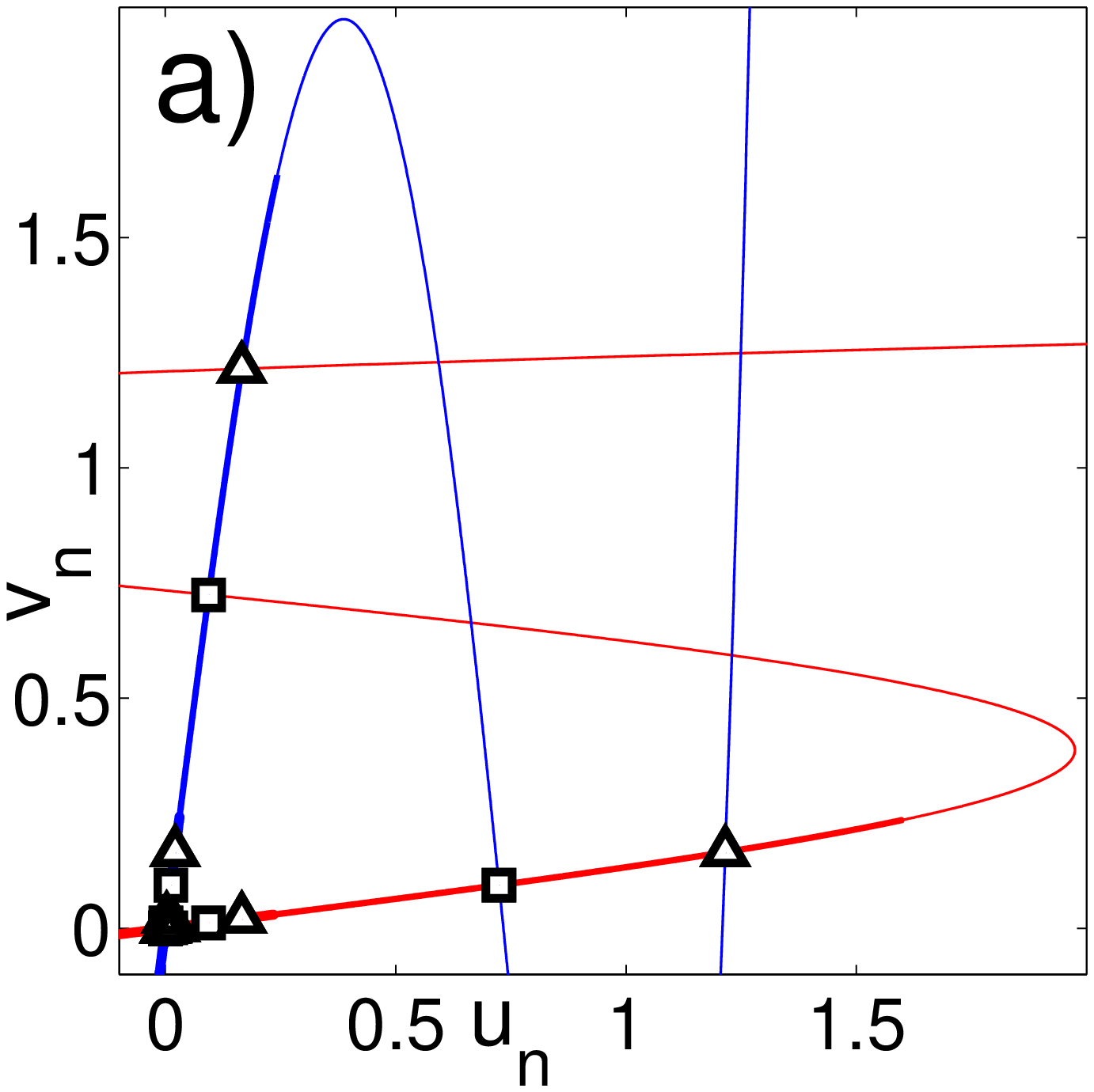,
width=5.3cm,angle=0,silent=} \epsfig{file=\rootfig
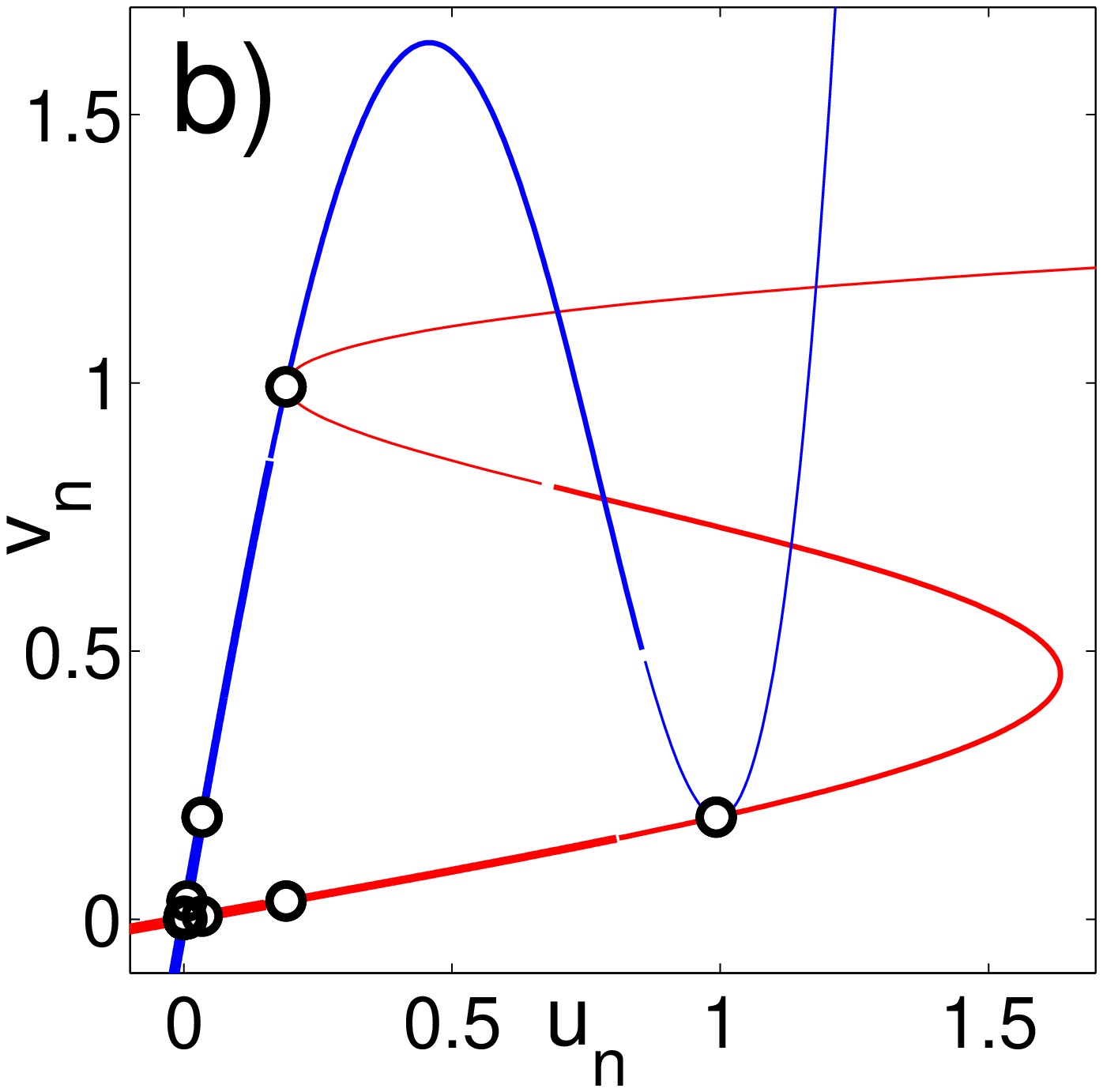, width=5.3cm,angle=0,silent=} }
\caption{Homoclinic tangles generating the solitons in Fig.\
\protect\ref{sols_mu06_mu07.ps}. a) the point $(\protect\mu
,C)=(-0.6,0.1)$ belongs to the bistability region (see asterisk in
Fig.\ \protect\ref{stability.ps}) that generates two solitons
(triangles, $T_{1}$, and squares, $S_{1}$). b) the point
$(\protect\mu ,C)=(-0.7,0.1855)$ is located at the bifurcation
curve (see the circle in Fig.\ \protect\ref{stability.ps}),
hence the two solutions collapse into one (see the right panel in
Fig.\ \protect\ref{sols_mu06_mu07.ps}). } \label{tangles.ps}
\end{figure}

\section{Bifurcations of the discrete solitons \label{Sec:bif}}

In this section we aim to explain how the $T_{1}$ solutions
disappear and how the evolution of the $S$ solutions leads to
stability changes and creation of pairs of \emph{stable
asymmetric} solitons. It is straightforward to understand the
simultaneous disappearance of the $S$ and $T$-type solutions as
$C$ increases. The boundary on which this happens corresponds to
the left side of the tent-like stability area of the $T$ solutions
in Fig.\ \ref{stability.ps}. The $S_{k}$ and $T_{k}$ solutions
collide on this boundary and disappear in a saddle-node
bifurcation. This bifurcation can be easily followed using the
homoclinic-tangle approach (see Fig.\ \ref{tangles.ps}). Before
the bifurcation occurs, i.e.,\ below the boundary (see the
asterisk in Fig.\ \ref{stability.ps}), two intersections,
identified by squares and triangles in Fig.\ \ref{tangles.ps}(a),
generate the solitons of the types $S_{1}$ and $T_{1}$,
respectively, which are are shown in the left panels of Fig.\
\ref{sols_mu06_mu07.ps}. In contrast, when the bifurcation curve
is approached, see Fig.\ \ref{tangles.ps}(b), the stable and
unstable manifolds barely intersect, and the two solitons are
nearly identical. Exactly at the bifurcation, the manifolds touch
tangentially, and only one solution is generated (i.e.,\ the
$S_{1}$ and $T_{1}$ solitons are identical at this point). After
this saddle-node bifurcation, these solutions do not exist
anymore.

\begin{figure}[th]
\centerline{ \epsfig{file=\rootfig 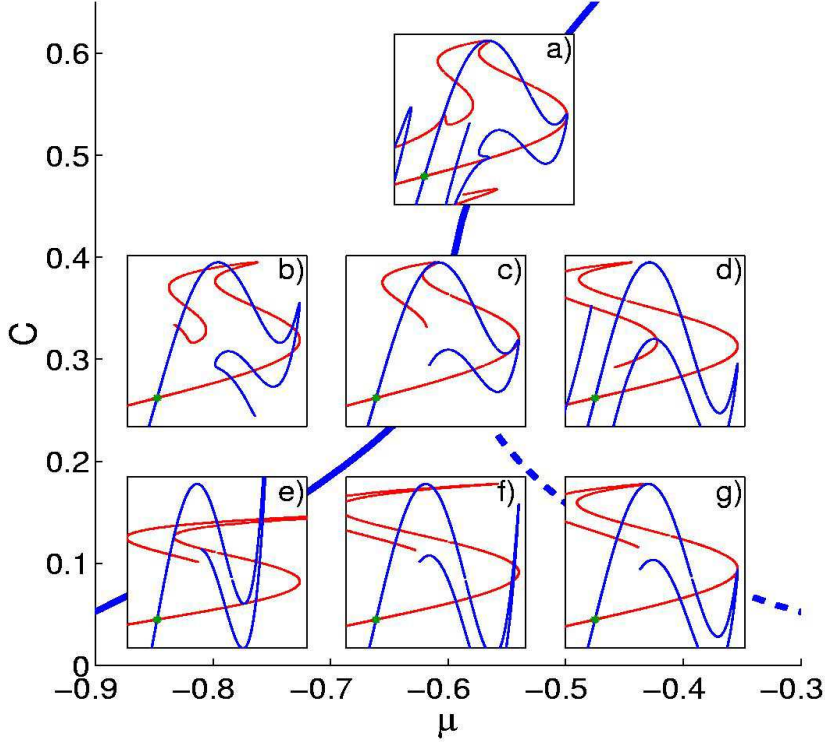, width=11cm,
height=10.0cm,angle=0,silent=} } \caption{Different types of
bifurcations around the cusp of the $T_{1}$ region. A saddle-node
bifurcation corresponding to panel (e) annihilates the pair of the
$S_{1}$ and $T_{1}$ solutions (the same bifurcation was depicted
in Fig.\ \protect\ref{tangles.ps}). Another saddle-node
bifurcation corresponds to panel (g) and is responsible for the
destruction of the $T_{1} $ solution. Going from (d) to (a), there
is a pitchfork bifurcation which is responsible for the loss of
stability of the symmetric soliton and the creation of a pair of
stable asymmetric ones (see further details in Fig.\
\protect\ref{inset_bif.ps}). } \label{inset.ps}
\end{figure}

Another family of noteworthy bifurcations points corresponds to
cuspidal points of the $T$-stability regions. Three types of
bifurcations occur near these points: (A) the saddle-node
bifurcation described above, at which the $S$ and $T$ solitons
collide and disappear; (B) another saddle-node bifurcation, at
which the $T$ solutions disappear; and (C) a bifurcation at which
the $S$ solitons lose their stability. These bifurcations are
depicted in Fig.\ \ref{inset.ps} as follows: (A) corresponds to
going through panels (f)$\rightarrow$(e)$\rightarrow$(b), (B)
is displayed by a chain of panels (f)$\rightarrow$(g)$\rightarrow$(d),
and (C) corresponds to going through
(d)$\rightarrow$(a)$\rightarrow$(b).

\begin{figure}[th]
\centerline{
\epsfig{file=\rootfig 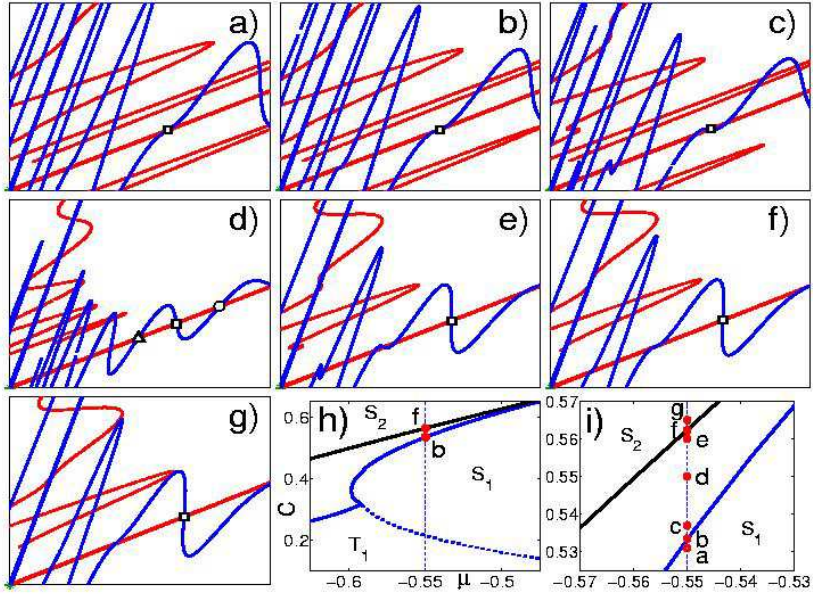, width=11cm, angle=0,silent=}
}
\caption{The pitchfork bifurcation leading to asymmetric soliton solutions.
This series of plots was obtained by keeping $\protect\mu =-0.55$ and
varying $C$ from the $S_{1}$ to $S_{2}$ regions, as shown in insets (h) and
(i). The series includes the following panels: (a) the $S_{1}$ region before
the bifurcation; (b) at the pitchfork bifurcation; (c), (d) and (e) in the
region where asymmetric solitons exist; (f) is a \textit{reverse pitchfork}
that destroys the asymmetric solutions; and (g) the $S_{2}$ region after the
latter bifurcation. The square dot represents the same homoclinic orbit
throughout the bifurcation, while the triangle and circle in (d) correspond
to the asymmetric solitons from Fig.\ \protect\ref{manyorbits}. }
\label{inset_bif.ps}
\end{figure}

Bifurcation (C) deserves more attention. In Fig.\ \ref{inset_bif.ps} we
depict in more detail the bifurcation scenario corresponding to a route from
the stability region of $S_{1}$ to its $S_{2}$ counterpart, which passes
through a gap where no $S$ solution is stable. Panels (a) through (g) show
the homoclinic tangles as $C$ increases, for fixed $\mu =-0.55$. Parameter
values for each of these panels are indicated in panel (h) and its
magnification, panel (i). From this figure, the existence of a pitchfork loop
---a supercritical pitchfork shortly followed by a reverse subcritical one--- is
evident. In panel (a) we depict by a square point one of the homoclinic
intersections, which gives rise to solution $S_{1}$. As $C$ increases, the
supercritical pitchfork bifurcation is reached, panel (b), that gives rise
to a pair of extra homoclinic intersections (see panel (c) and
triangle and circle in
panel (d)). As $C$ increases further, the reverse subcritical pitchfork
occurs, eliminating the extra pair of intersections.

Quite interesting are the shape and stability of solitons that
this extra pair generates in the narrow gap between the $S_{k}$
and $S_{k+1}$ stability regions (which coincides with the region
of the pitchfork loop), where these new solutions exist. They
represent \emph{asymmetric} solitons of the CQ DNLS equation,
which are depicted in Fig.\ \ref{manyorbits}(b) by means of
triangles and circles. As might be expected, the stability is
swapped by the pitchfork bifurcations. Indeed, the stable $S$
solitons are unstable inside the pitchfork loop, where the new
asymmetric solutions are \emph{stable}. An example of unstable
evolution of an $S$ soliton inside the pitchfork loop is shown in
the bottom row of Fig.\ \ref{unstable_evol.ps}. We stress that
such pitchfork loops, and the respective asymmetric solitons, do
not exist in the cubic DNLS equation (asymmetric solitons were
also found in a DNLS equation with a mixture of cubic onsite and
intersite nonlinearities \cite{Johansson}).

Further inspection of the pitchfork loop demonstrates that, if the
$S_{1}$ \emph{site-centered} soliton looses its stability through
the direct bifurcation, then the reverse bifurcation, which closes
the loop, stabilizes an $S_{2}$ \emph{bond-centered} soliton. At
the next pitchfork loop (in the gap between stability of $S_{2}$
and $S_{3}$), the latter bond-centered solution gets destabilized
and, after the loop, an $S_{3}$ \emph{site-centered} soliton gains
its stability. As $C$ is increased further, this process repeats
itself, creating stability bands for the $S_{k}$ solitons for
larger $k$. In fact, all the solutions $S_{k}$ represent only two
distinct soliton families emanating from the principal homoclinic
intersection: of site-centered solutions, and of bond-centered
ones, the pitchfork loops conducting stability swaps between the
two families. For given $S_{k}$, the number of the pitchfork loops
passed by the soliton, while developing from the anti-continuum
($C=0$) limit, is $k-1$.

The existence of \emph{stable} bond-centered solitons is a noteworthy
feature by itself, as in the usual DNLS model with the cubic nonlinearity
only site-centered states are stable (in the DNLS equation with the
saturable onsite nonlinearity, stable bond-centered solitons were found too
\cite{discrete-saturable}). An interesting issue which still has to be
addressed is whether there exists an accumulation curve for the pitchfork
loops, or the loops continue to appear as one approaches the continuum
limit, $C=\infty $.

\section{The variational approximation \label{Sec:VA}}

The variational approximation (VA) can be used to describe the
shape of stationary soliton solutions in an analytical form. The
method will not only be successful in approximating the shape of
the most fundamental solitons, but will also predict the
saddle-node bifurcation where the $T_{1}$ and $S_{1}$ solutions
collide and disappear.

The only tractable \emph{ansatz} for the VA in the discrete models is one
based on an exponential cusp, that was applied in Ref.\ \cite{Weinstein} to
solitons in the above-mentioned DNLS equation with an arbitrary power
nonlinearity :
\begin{equation}
u_{n}=Ae^{-\alpha |n|},  \label{ansatz}
\end{equation}
with real positive constants $A$ and $\alpha $. We identify the value of
$\alpha $ without the resort to the VA proper, but rather from the
substitution of ansatz (\ref{ansatz}) in Eq.\ (\ref{CQDNLS-sta}) linearized
for the decaying tail of the soliton, which yields
\begin{equation}
\alpha =\ln \left( \frac{a}{2}+\sqrt{\left( \frac{a}{2}\right) ^{2}-1}\right)
\label{eigen}
\end{equation}(recall $a\equiv 2-\mu /C$). Thus, a necessary (but not sufficient)
condition for the existence of any soliton solution is $a>2$, or $\mu <0$
(since we set $C>0$, it will then guarantee that Eq. (\ref{eigen}) yields a
real positive $\alpha $). We stress that the latter soliton-existence
condition is an \emph{exact} one, as it follows from the straightforward
consideration of the exponentially localized tails of the soliton, and does
not exploit any approximation.

Now we invoke the VA proper, treating the amplitude $A$ in ansatz
(\ref{ansatz}) as a variational parameter, while $\alpha $ was
already fixed as per Eq. (\ref{eigen}) (i.e., by the condition
that the ansatz must match the correct asymptotic form of the
soliton solution).

\begin{figure}[th]
\centerline{ \epsfig{file=\rootfig 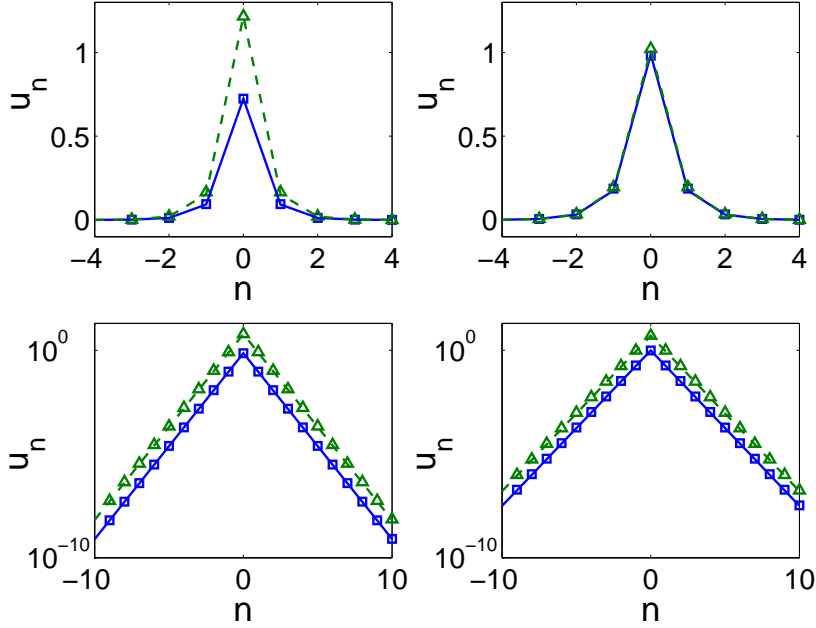, width=11cm,
angle=0,silent=} } \caption{Soliton solutions of Eq.\
(\protect\ref{CQDNLS-sta}) for $(\protect\mu ,C)=(-0.6,0.1)$
(left) and $(\protect\mu ,C)=(-0.7,0.1845)$ (right) are shown on
straight (top) and logarithmic (bottom) scales. Points and lines
correspond, respectively, to numerically exact solutions and ones
predicted by the variational approximation. The right panels
correspond to a point $(\protect\mu ,C)$ close to the collision of
the soliton solutions of the $S_{1}$ and $T_{1}$ types (see the
circle in Fig.\ \protect\ref{stability.ps}).}
\label{sols_mu06_mu07.ps}
\end{figure}

The stationary equation\ (\ref{CQDNLS-sta}) can be derived from the Lagrangian
\begin{equation}
{L}=\displaystyle\sum_{n=-\infty }^{\infty }\left[ \mu
u_{n}^{2}+\frac{B}{2}u_{n}^{4}-\frac{Q}{3}u_{n}^{6}-C(u_{n+1}-u_{n})^{2}\right]
.  \label{Lag2}
\end{equation}
The substitution of the ansatz into the Lagrangian and explicit calculation
of the sum lead to the following effective Lagrangian, which is the main
ingredient of the VA\ \cite{borisVA}:
\begin{equation}
{L}_{\mathrm{eff}}{(A)}=\displaystyle\mu A^{2}\coth (\alpha )+A^{4}\coth
(2\alpha )-\frac{A^{6}}{3}\coth (3\alpha )-2\tanh (\alpha /2)CA^{2}.
\label{LagA}
\end{equation}Then, the Euler-Lagrange equation, $d{L}_{\mathrm{eff}}{(A)/dA=0}$, yields a
quadratic equation for $A^{2}$,
\begin{equation}
-\coth (3\alpha )\left( A^{2}\right) ^{2}+2\coth (2\alpha )\left(
A^{2}\right) +\mu \coth (\alpha )-2\tanh (\alpha /2)C=0.  \label{quad}
\end{equation}

\begin{figure}[th]
\centerline{
\epsfig{file=\rootfig 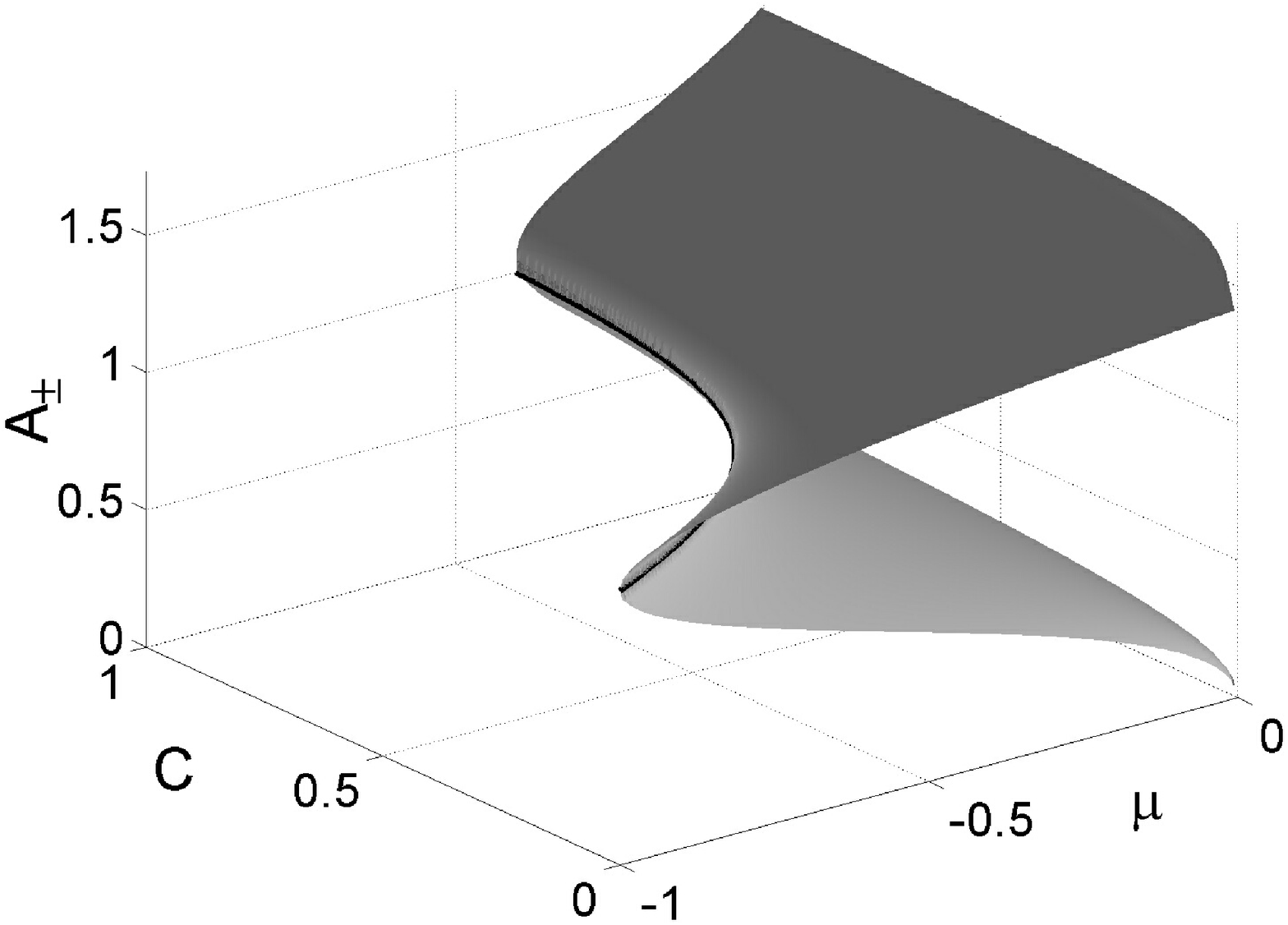, width=6.7cm,angle=0,silent=}
~~
\epsfig{file=\rootfig 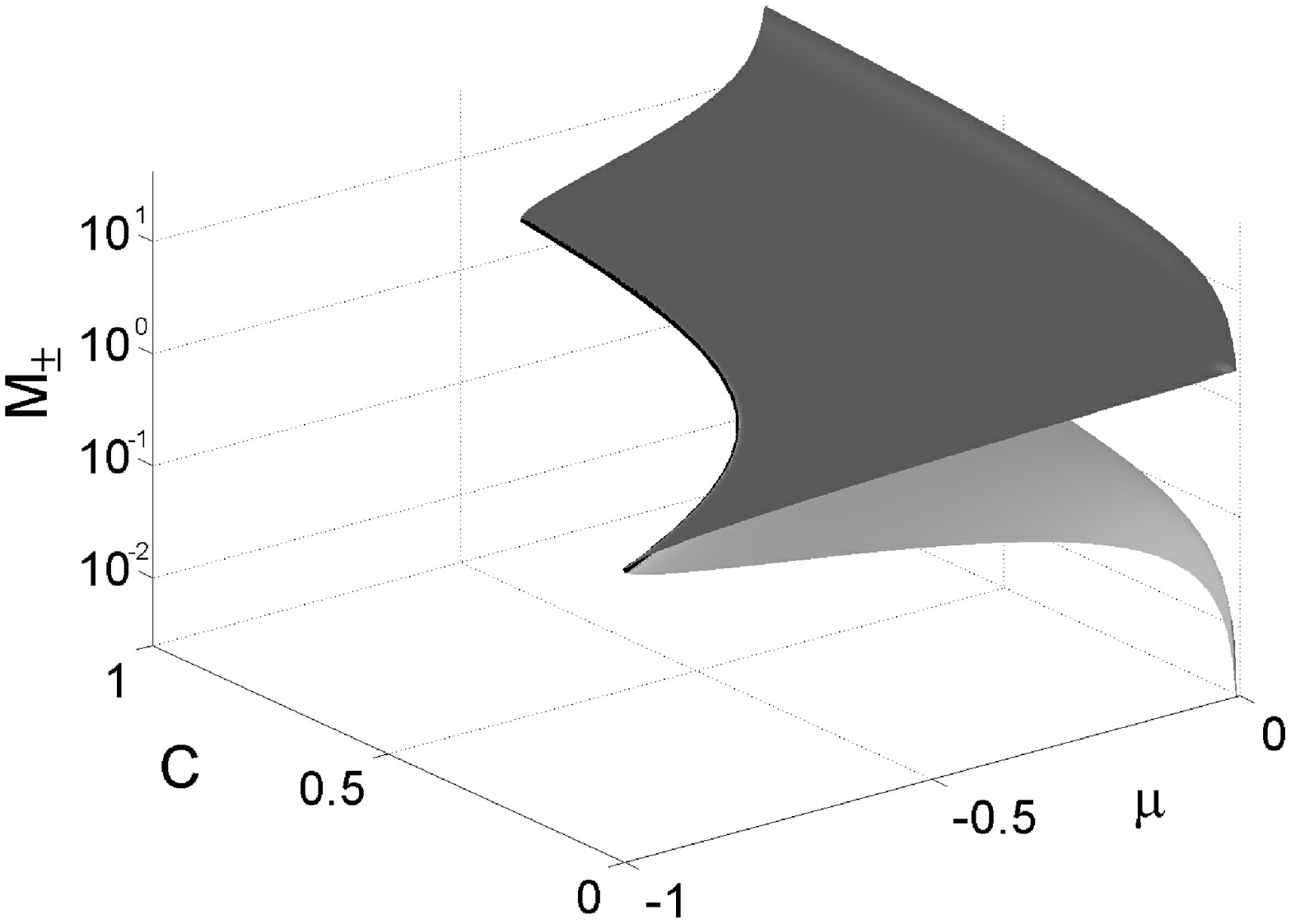, width=6.7cm,angle=0,silent=}
}
\caption{Amplitude $A$ (left) and norm $M$ (right) for the two soliton
solutions predicted by the VA (the scale for the norm is logarithmic). The
top and bottom branches correspond to the tall and short solitons,
respectively. The dark solid line represent the bifurcation curve where the
two solutions are predicted to coalesce and disappear.}
\label{Ap_Am.ps}
\end{figure}

Equation\ (\ref{quad}) indicates that there may be two different solutions.
The top panel of Fig.\ \ref{Ap_Am.ps} depicts the amplitude associated
with these two solutions predicted by the VA. It is clear that the
two solutions coalesce and disappear at the bifurcation curve (solid line).
In Fig.\ \ref{sols_mu06_mu07.ps} we display examples of these
two variational solutions, together with the ones obtained numerically from
Eq.\ (\ref{CQDNLS-sta}) through Newton iterations. It is observed that the
match between the numerical and variational solutions is extremely good for
small values of the coupling parameter $C$. For larger values of $C$, the
solution tends to its continuum (smooth) analogue where the cusp-shaped
ansatz (\ref{ansatz}) is obviously irrelevant.

The VA predicts the bistable solutions of the type (\ref{ansatz})
inside the $(\mu ,C)$-parameter regions where the quadratic
equation (\ref{quad}) admits two distinct positive roots. The
comparison with numerical results in Fig.\ \ref{sols_mu06_mu07.ps}
immediately shows that the two different variational solutions
exactly correspond to the $S_{1}$ and $T_{1}$ solitons, as they
were defined above. Therefore, the curve where Eq.\ (\ref{quad})
has a double root represents the saddle-node collision of $S_{1}$
and $T_{1}$. This bifurcation curve predicted by the VA is
depicted by the dashed-dotted curve labeled VA1 in Fig.\
\ref{stability.ps}. It is quite remarkable that the approximation
based on the simple ansatz (\ref{ansatz}) is able to capture the
saddle-node bifurcation so well for small $C$.

It is possible to refine the VA approach by allowing a more general ansatz
of the form
\begin{equation}
u_{n}=\left\{
\begin{array}{lcl}
\phantom{\beta}Ae^{-\alpha |n|} & \mathrm{~if~} & |n|>0, \\[0.5ex]
\beta A & \mathrm{~if~} & n=0,\end{array}\right.
\label{ansatzVA2}
\end{equation}
where we introduce a new free parameter $\beta $. This new ansatz is able to
predict the existence of the solitons of two distinct types, which can me
immediately identified with short ($S$) and tall ($T$) solitons, that were
described in detail above.

The new variational ansatz (\ref{ansatzVA2}), with two independent
variational parameters $(A,\beta )$, is able to very accurately
predict the whole existence region of the $T_{1}$ solution, as
well as the saddle-node bifurcation which creates the $S_{3}$
solution. The boundary of the $S_{3}$ solitons predicted by the
improved VA is depicted by the dash-dotted line VA2 in
Fig.~\ref{stability.ps}. As seen from the figure, the new VA gives
an extremely good approximation for the boundary, up to $C\lesssim
1$. The results for the existence region of the $T_{1}$ soliton
predicted by this VA are not depicted in Fig.~\ref{stability.ps}
because they exactly coincide (up to the resolution of the figure)
with \emph{both}, left and right, numerically found boundaries for
the $T_{1}$ soliton, see the darker shaded area in
Fig.~\ref{stability.ps} (the simple VA, based on ansatz
(\ref{ansatz}), which gives rise to curve VA1, only captured the
left boundary).

The improved VA with ansatz (\ref{ansatzVA2}) is also able to pick
up solutions with a dip at the central site, corresponding to
$\beta <e^{-\alpha}$ in the ansatz, i.e., a bound state of two
solitons, or multibreather~\cite{ref:multibreathers}).  Thus,
the modified VA is able to describe the respective bifurcation
curves too (results not presented here).

A well-known approach to predicting the stability for soliton
families is based on the Vakhitov-Kolokolov (VK) criterion
\cite{VK}. The VK criterion states that a solution family,
parameterized by the frequency $\mu $, as in Eq. (\ref{mu}), may
be stable if $dM/d\mu <0$, and is definitely unstable otherwise,
where $M$ is the soliton's norm defined as per Eq. (\ref{M}). The
simplest variational ansatz (\ref{ansatz}) yields $M=A^{2}\coth
(\alpha )$ [note that both $\alpha $ and $A$ depend on $(\mu ,C)$
through Eqs.\ (\ref{eigen}) and (\ref{quad})]. The norm given by
the latter expression is depicted, as a function of $(\mu ,C)$, in
the right panel of Fig.\ \ref{Ap_Am.ps}. It is clear from the
figure that $dM/d\mu >0$ for the $T_{1}$ soliton, and $dM/d\mu <0$
for $S_{1}$ (these results can be proven analytically, within the
framework of the VA). Thus, the VK criterion suggests that the
$T_{1}$ soliton is unstable, while its $S_{1}$ counterpart may be
stable. Comparing this conclusion with the numerical results for
the stability reported above we conclude that the VK criterion
\emph{does not} apply to our model. In fact, similar conclusions
for the failure of the VK criterion where obtained in the
\emph{continuous} CQ NLS equation with an external potential, that
might be both a single rectangular potential well
\cite{BorisKP:04}, and a periodic lattice of rectangular wells
(the above-mentioned Kronig-Penney potential) \cite{BorisKP:05}.

\begin{figure}[th]
\centerline{
\epsfig{file=\rootfig 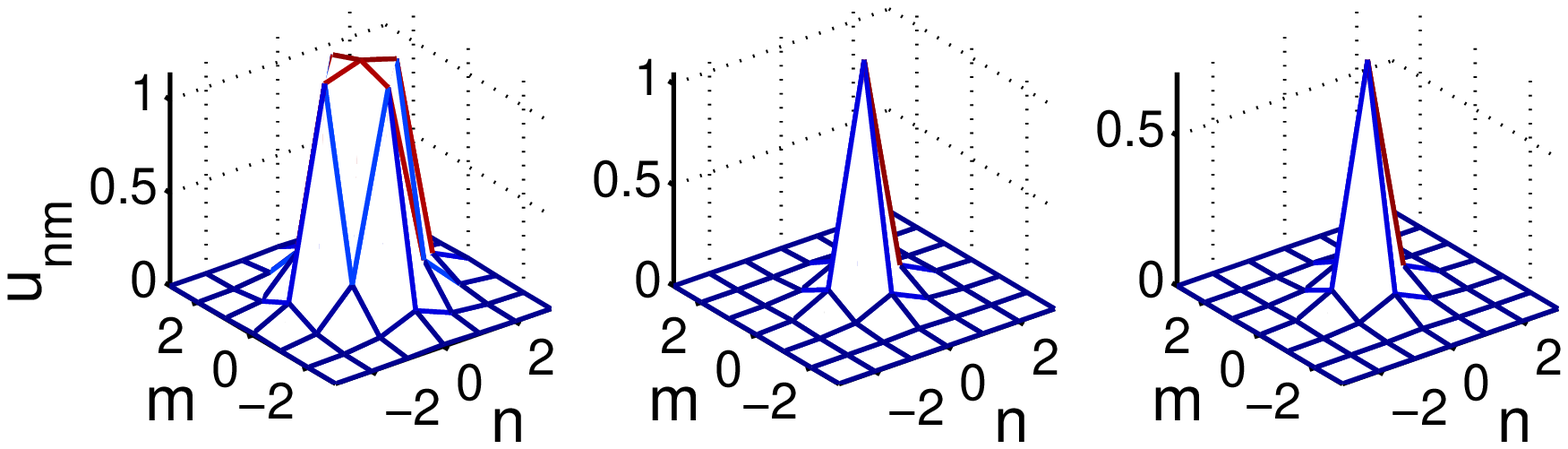, width=12cm, angle=0,silent=}
}
\centerline{
\epsfig{file=\rootfig 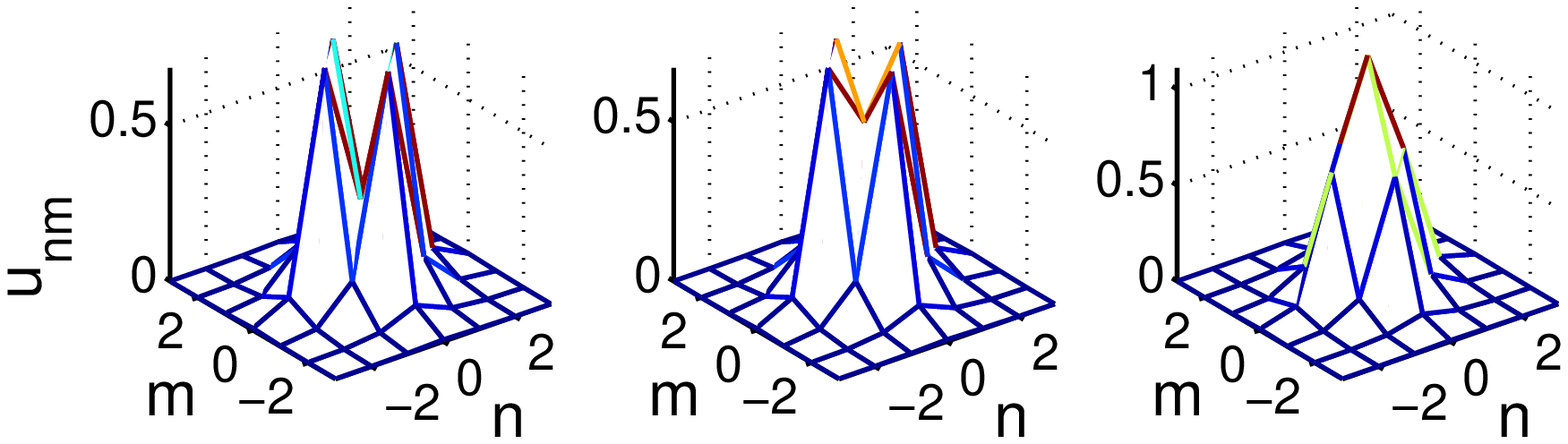, width=12cm, angle=0,silent=}
}
\caption{Two-dimensional coexisting soliton solutions for $(\mu,C)=(-0.6,0.07)$.
The top row depicts three coexisting {\em stable} soliton solutions
while the bottom row represents  three coexisting {\em unstable}
soliton solutions. All these solutions were obtained by
Newton iteration on the steady state equation with initial
seeds predicted by the two-dimensional variational approximation
with ansatz (\ref{ansatz2DVA2}).
}
\label{2DVA2.ps}
\end{figure}

A natural question that arises is the existence of multistable
solutions in higher dimensional lattices.
The higher dimensional equivalent of (\ref{CQDNLS})
is obtained by replacing the one-dimensional index $n$
and the discrete Laplacian by their higher-dimensional analogues.
Unfortunately, the homoclinic approach is not applicable
in higher dimensional lattices.
Nonetheless, the variational approach presented here is still
amenable in the higher dimensional case. Namely, a
natural ansatz equivalent to (\ref{ansatzVA2}) but now in
two dimensions (2D) becomes
\begin{equation}
u_{nm}=\left\{
\begin{array}{lcl}
\phantom{\beta}Ae^{-\alpha (|n|+|m|)} & \mathrm{~if~} & |n|+|m|>0, \\[0.5ex]
\beta A & \mathrm{~if~} & n=m=0,\end{array}\right.
\label{ansatz2DVA2}
\end{equation}
where now we have two indexes $(m,n)$ along the two spatial dimensions,
and the decay $\alpha$ is the same as for the one-dimensional case
[cf.\ Eq.\ (\ref{eigen})]. Preliminary results indicate the
coexistence of {\em stable} soliton solutions in the 2D model.
As an example we depict in Fig.\ \ref{2DVA2.ps} six coexisting
solutions for the same parameter values. The top row
in the figure depicts three {\em stable}
coexisting solutions. It is worth
mentioning that the variational approach for the 2D case is also
able to capture other solutions such as, stable and unstable,
multi-humped profiles (see bottom row of Fig.\ \ref{2DVA2.ps}).
A detailed stability analysis for higher
dimensional (2D and 3D) soliton solutions falls outside of
the scope of the present manuscript and will be presented
elsewhere.

\section{Conclusions and extensions\label{Sec:conclusions}}

The objective of this work was to introduce the discrete nonlinear
Schr\"{o}dinger (DNLS) equation with the competing cubic-quintic
(CQ) nonlinearities. Besides being a new dynamical model, this
model may also apply to the description of an array of optical
waveguides with the intrinsic CQ nonlinearity.

We have studied the multistability of discrete solitons in the
model, looking for homoclinic solutions to the respective
stationary discrete equation. Regions of the existence and
stability of single-humped soliton solutions were identified by
using a numerical continuation method based on Newton-type
iterations. The stability of the various types of the solitons
was investigated through numerical evaluation of eigenvalues for
small perturbations. The resulting stability diagram suggests the
existence of an infinite family of branches of \emph{stable}
solitons of the aforementioned type $T_{k}$, for sufficiently
small values of the coupling constant $C$. As $C$ increases, these
solutions get destroyed through saddle-node bifurcations. We have
also identified another type of discrete soliton solutions,
$S_{k}$, with $S_{1}$ surviving in the continuum limit,
$C\rightarrow \infty $. The stability of solutions of the latter
type ($S$) changes, with the increase of $C$, through a series of
small pitchfork loops (each opens with a supercritical pitchfork,
which is shortly followed by a reverse supercritical one). Inside
the loops the symmetric solitons $S_{k}$ loose their stability,
while a pair of \emph{stable asymmetric} solutions is created. The
latter solutions have no counterpart in the cubic DNLS equation,
nor in the continuum CQ\ NLS equation. Finally, using the
variational approximation (VA), we were able to approximate the
main branches of the solutions and their bifurcations for small
$C$. We have also proposed an improved version of the VA, with two
free parameters rather than one, that drastically (qualitatively)
upgrades the accuracy of the VA, making it possible to predict
simultaneously the most fundamental solitons and some higher-order
ones.

An interesting extension of this work would be to thoroughly
investigate localized solutions
in the two-dimensional version of the CQ DNLS equation, which may be
realized, for instance, as a \emph{bunch} of the corresponding nonlinear
optical waveguides (cf. the recently reported experimental realization for
linear waveguides \cite{Jena}). In particular, the existence and stability
of \emph{asymmetric} solitons in the two-dimensional model would be an issue
of great interest. Another promising direction for further considerations
may be the study of kink solutions, corresponding to heteroclinic
trajectories generated by the intersection of the stable and unstable
manifolds of different fixed points of the map (\ref{2Dmap}).

As concerns the underlying mathematical theory, an intriguing issue is to
understand why the Vakhitov-Kolokolov criterion was found to consistently
fail in the discrete CQ NLS equation and its counterpart with a periodic
potential in the continuum case \cite{BorisKP:05,BorisKP:04}.

\section*{Acknowledgments}

We appreciate valuable discussions with J.\ Fujioka and A.\
Espinosa. B.A.M.\ appreciates the hospitality of the Institute of
Physics at the Universidad Nacional Aut\'{o}noma de M\'{e}xico,
and the Nonlinear Dynamical Systems
group\footnote{http://nlds.sdsu.edu} at the Department of
Mathematics and Statistics, San Diego State University. This
author was partially supported by a Window-on-Science grant
provided by the European Office of Aerospace Research and
Development of the US Air Force. R.C.G.\ and C.C.\ acknowledge the
Grant-in-Aid award provided by SDSU Foundation. R.C.G.\
also acknowledges support from NSF-DMS-0505663. J.D.T.\ is
grateful for the support of the Computational Science Research
Center\footnote{http://www.sci.sdsu.edu/csrc/} at SDSU.

\end{document}